\documentclass[twocolumn,showpacs,printnumbers,amsmath,amssymb,pra]{revtex4}

\usepackage{bm}
\usepackage[dvipdfm]{graphicx}
\usepackage[
  dvipdfm,
  bookmarks=true,
  bookmarksnumbered=true,
  bookmarkstype=toc,
  pdftitle={},
  pdfsubject={},
  pdfauthor={Motoaki BAMBA}
]{hyperref}
\usepackage{color}

\newcommand{\bra}[1]{\langle #1|}
\newcommand{\ket}[1]{|#1\rangle}
\newcommand{\braket}[1]{\langle #1 \rangle}
\newcommand{\braketin}[1]{\langle #1 \rangle_{\text{in}}}

\newcommand{\braketg}[1]{\langle #1 \rangle_{\text{g}}}

\newcommand{\braketR}[1]{\langle #1 \rangle_{\text{R}}}

\def\dd{\mathrm{d}}
\def\ee{\mathrm{e}}
\def\ii{\mathrm{i}}

\def\ddt#1{\frac{\partial #1}{\partial t}}
\def\ddtau#1{\frac{\partial #1}{\partial\tau}}
\def\PP{\mathrm{P}}
\def\Tr{\mathrm{Tr}}
\def\const{\text{const.}}
\def\Hc{\text{H.c.}}

\def\low{\downarrow}
\def\rai{\uparrow}

\def\ketg{\ket{\text{g}}}
\def\brag{\bra{\text{g}}}

\def\ketgg{\ket{\tilde{\text{g}}}}
\def\bragg{\bra{\tilde{\text{g}}}}

\def\wc{\omega_{\text{c}}}
\def\wx{\omega_{\text{x}}}

\def\wn{\tilde{\omega}}
\def\rabi{g}
\def\AA{D}
\def\wco{\varOmega^{\text{cut-off}}}

\def\DOSc{n_{\text{DOS}}^c}
\def\DOSx{n_{\text{DOS}}^x}

\def\Gc{G_c}
\def\Gx{G_x}

\def\ww{\tilde{w}}

\def\oH{\hat{H}}
\def\oHz{\hat{H}_{\text{S}}}
\def\oHint{\hat{H}_{\text{S-R}}}
\def\oHR{\hat{H}_{\text{R}}}
\def\oHa{\hat{H}_{\text{ph}}}
\def\oHb{\hat{H}_{\text{ex}}}

\def\orho{\hat{\rho}}
\def\orhoss{\hat{\rho}_{\text{ss}}}
\def\orhotot{\hat{\rho}_{\text{tot}}}
\def\orhoR{\hat{\rho}_{\text{R}}}

\def\oa{\hat{a}}
\def\oad{\hat{a}^{\dagger}}
\def\ob{\hat{b}}
\def\obd{\hat{b}^{\dagger}}

\def\oO{\hat{O}}
\def\oS{\hat{S}}

\def\os{\hat{s}}
\def\osd{\hat{s}^{\dagger}}

\def\ooa{\hat{A}}
\def\ooad{\hat{A}^{\dagger}}
\def\oob{\hat{B}}
\def\oobd{\hat{B}^{\dagger}}
\def\oop{\hat{P}}
\def\oopd{\hat{P}^{\dagger}}
\def\uc{u_c}
\def\ux{u_x}
\def\vc{v_c}
\def\vx{v_x}
\def\ww{W}
\def\xx{X}
\def\yy{Y}
\def\zz{Z}
\def\winv{\omega'}

\def\oalpha{\hat{\alpha}}
\def\oalphad{\hat{\alpha}^{\dagger}}

\def\oalphain{\hat{\alpha}^{\text{in}}}
\def\oalphaout{\hat{\alpha}^{\text{out}}}
\def\obeta{\hat{\beta}}
\def\obetad{\hat{\beta}^{\dagger}}

\def\obetain{\hat{\beta}^{\text{in}}}

\def\op{\hat{p}}
\def\opd{\hat{p}^{\dagger}}

\def\oFnz{\check{F}^{(0)}}
\def\oFnzd{\check{F}^{(0)\dagger}}
\def\ovFnz{\check{\bm{F}}^{(0)}_{LU}}

\def\oF{\hat{F}}
\def\oFd{\hat{F}^{\dagger}}
\def\oFz{\hat{F}^{(0)}}
\def\oFzd{\hat{F}^{(0)\dagger}}

\def\oFin{\hat{F}^{\text{in}}}
\def\oFind{\hat{F}^{\text{in}\dagger}}
\def\oFout{\hat{F}^{\text{out}}}
\def\oFoutd{\hat{F}^{\text{out}\dagger}}

\def\ovFin{\hat{\bm{F}}^{\text{in}}}

\def\oFfwd{\hat{F}^{\text{fwd}}}

\def\oAL{\hat{\mathcal{A}}^{\text{L}}}
\def\oAR{\hat{\mathcal{A}}^{\text{R}}}
\def\oBL{\hat{\mathcal{B}}^{\text{L}}}
\def\oBR{\hat{\mathcal{B}}^{\text{R}}}
\def\oCL{\hat{\mathcal{C}}^{\text{L}}}
\def\oCR{\hat{\mathcal{C}}^{\text{R}}}
\def\oDL{\hat{\mathcal{D}}^{\text{L}}}
\def\oDR{\hat{\mathcal{D}}^{\text{R}}}

\def\oL{\hat{\mathcal{L}}}
\def\oLdiss{\hat{\mathcal{L}}_{\text{diss}}}
\def\oU{\hat{U}}
\def\oUz{\hat{U}_{\text{S}}}

\def\mL{\bm{\mathsf{L}}}
\def\mM{\bm{\mathsf{M}}}
\def\mV{\bm{\mathsf{V}}}
\def\munit{\bm{1}}

\begin{document}
\title{Dissipation and detection of polaritons in ultrastrong coupling regime}
\author{Motoaki Bamba}
\altaffiliation{E-mail: bamba@acty.phys.sci.osaka-u.ac.jp}
\affiliation{Department of Physics, Osaka University, 1-1 Machikaneyama, Toyonaka, Osaka 560-0043, Japan}
\author{Tetsuo Ogawa}
\affiliation{Department of Physics, Osaka University, 1-1 Machikaneyama, Toyonaka, Osaka 560-0043, Japan}
\date{\today}

\begin{abstract}
We have investigated theoretically a dissipative polariton system
in the ultrastrong light-matter coupling regime
without using the rotating-wave approximation
on system-reservoir coupling.
Photons in a cavity and excitations in matter respectively couple
two large ensembles of harmonic oscillators (photonic and excitonic reservoirs).
Inheriting the quantum statistics of polaritons in the ultrastrong coupling regime,
in the ground state of the whole system,
the two reservoirs are not in the vacuum states
but they are squeezed and correlated.
We suppose this non-vacuum reservoir state in 
the master equation and in the input-output formalism with Langevin equations.
Both two approaches consistently guarantee
the decay of polariton system to its ground state,
and no photon detection is also obtained 
when the polariton system is in the ground state.
\end{abstract}


\maketitle

\section{Introduction}
Light-matter ultrastrong coupling \cite{Ciuti2005PRB,Ciuti2006PRA,Devoret2007AP,Bourassa2009PRA,DeLiberato2009PRA,Gunter2009N,Anappara2009PRB,Todorov2009PRL,Todorov2010PRL,Hagenmuller2010PRB,Nataf2010PRL,Nataf2010NC,Peropadre2010PRL,Niemczyk2010NP,Forn-Diaz2010PRL,Fedorov2010PRL,Ashhab2010PRA,Meaney2010PRA,Casanova2010PRL,Hausinger2010PRA,Hausinger2011PRA,Nataf2011PRL,Schwartz2011PRL,Beaudoin2011PRA,Scalari2012S,Todorov2012PRB,Porer2012PRB,Ridolfo2012arXiv}
means that the coupling strength $\rabi$ is comparable to or larger than the transition
frequency $\wx$ of excitations in matters ($\rabi \gtrsim \wx$),
and it shows a variety of peculiar properties,
such as virtual photons in the ground state \cite{Ciuti2005PRB,DeLiberato2009PRA},
squeezed eigen states \cite{Ciuti2005PRB,Casanova2010PRL},
nearly degenerated ground states \cite{Nataf2010PRL,Ashhab2010PRA,Meaney2010PRA},
quantum phase transitions \cite{Emary2004PRA,Nataf2010NC},
and so on.
The ultrastrong coupling has been realized experimentally by
intersubband transitions in semiconductor quantum wells \cite{Gunter2009N,Anappara2009PRB,Todorov2009PRL,Todorov2010PRL,Todorov2012PRB,Porer2012PRB},
artificial atoms in superconducting circuits \cite{Niemczyk2010NP,Fedorov2010PRL,Forn-Diaz2010PRL},
and cyclotron transition in two-dimensional electron gas \cite{Scalari2012S}.
In most cases, photons with THz or microwave frequency
are confined in a cavity,
and the cavity photons are coupled with external photonic field (outside the cavity)
by unignorable dissipation rate
(but small compared to the light-matter coupling).
In other words,
when we neglect the coupling with matters,
the real eigen modes of photons are represented as
coupled fields of cavity mode and external fields
\cite{Dutra2000JOB}.
As we will discuss in the present paper
by the Fano-type diagonalization technique
\cite{Fano1961PR,Barnett1988OC,huttner92},
when the cavity mode is squeezed in the ground state
due to the ultrastrong light-matter coupling,
the external photonic field is also squeezed
in the ground state of the whole system.
However, we cannot observe the squeezing nor energy flow
by photon detectors if the system is in the ground state.
In this paper, we have developed two frameworks,
the master equation and Langevin equations with input-output relation.
They are derived by supposing the squeezed external fields
and show no photon detection consistently.

In the standard theory of quantum optics \cite{Carmichael1973JPA,Carmichael1987JOSAB,gardiner04,Breuer2006,walls08,Ciuti2006PRA,Scala2007PRA,Scala2007JPA,Fleming2010JPA,Nakatani2010JPSJ,Beaudoin2011PRA},
in order to introduce a dissipation of cavity mode or of excitations in matters,
we consider coupling with an ensemble of harmonic oscillators
with continuous frequencies,
and the oscillators are supposed to be in the vacuum state.
The dissipation of focusing system has been successfully described
by such a treatment in both master equation and input-output formalism,
at least in the weak and (normally) strong light-matter coupling regimes
($\rabi \lesssim \varGamma$ and $\varGamma \lesssim \rabi \ll \wx$, respectively,
for dissipation rate $\varGamma$).
As pointed out in some papers
\cite{Carmichael1973JPA,Carmichael1974PRA,Scala2007PRA,Scala2007JPA,Fleming2010JPA,Nakatani2010JPSJ},
the master equation should be derived by considering
the eigen states of the focusing system,
and the rotating-wave approximation (RWA)
should be performed carefully on the system-reservoir coupling
even if the system-reservoir coupling is weak
compared to the light-matter coupling (in the strong light-matter coupling regime).
In the ultrastrong coupling regime, such treatment has been performed
by Beaudoin, Gambetta, and Blais \cite{Beaudoin2011PRA}.
The dissipation of the ultrastrong coupling systems
can be successfully described
under the RWA on system-reservoir coupling
(both pre-trace and post-trace RWAs are used in terms of Ref.~\cite{Fleming2010JPA})
by considering the eigen states of the cavity system
(there are squeezed virtual photons in the ground state).
Furthermore, the photon detection has also been discussed
in Ref.~\cite{Ridolfo2012arXiv}, and virtual photons in the cavity are not counted
by normal- and time-ordering the operators of cavity system.
However, the RWA on the system-reservoir coupling destroys the information on
how the cavity system couples with the reservoirs.
In other words, the reservoirs become a black box,
and we effectively suppose simplified reservoirs coupling with the eigen states.
Although such a treatment is appropriate and simple for discussing the dynamics
of the focusing system,
it becomes difficult to discuss the statistics of output photons emitted
from the cavity.
There are actually virtual photons not only in the cavity but also
in the photonic reservoir
even if the whole system is in the ground state,
which is naturally derived by the Fano-type diagonalization,
although the virtual photons are not counted by detectors.
Of course, when the cavity system is excited,
we can detect photons emitted from the cavity.
Whereas antibunching of emission can survive
even under the RWA on system-reservoir coupling \cite{Ridolfo2012arXiv},
the quantum fluctuation (squeezing) of the emission is easily diminished
in such treatment,
because the interference between reservoir free field and cavity contribution
is important for squeezing \cite{Carmichael1987JOSAB,gardiner04,walls08}.
Therefore, in order to fully discuss the quantum statistics of emission
in the ultrastrong light-matter coupling regime,
we have to develop a comprehensive framework
describing the dissipation and emission
without using the RWA on light-matter coupling
nor on system-reservoir one.

In the weak and normally strong light-matter coupling regimes,
the ground state of the whole system is approximately represented
by the vacuum states of photonic and excitonic reservoirs
(exactly the vacuum states under the RWA on light-matter coupling).
In both master equation and input-output formalism,
the vacuum reservoirs are usually considered for describing the dissipation,
and no photon detection is naturally obtained in the ground state.
Then, these three approaches (analysis of ground state, master equation,
and input-output formalism)
are consistent in the standard dissipation theory
in the weak and strong light-matter coupling regimes.
However, if we simply suppose the vacuum reservoirs,
in the ultrastrong light-matter coupling regime,
the master equation and input-output formalism give different results.
This is because the photonic and excitonic reservoirs
are not in the vacuum states in the ground state of the whole system,
but they are actually squeezed and correlated.
We have to suppose this non-vacuum reservoir state
in order to remove the discrepancy
between the results of master equation and of input-output formalism.

In the present paper, we discuss polariton system consisting of two bosonic modes,
photons in a cavity and excitations in matter, each of which couples with 
an ensemble of harmonic oscillators (photonic and excitonic reservoirs).
Diagonalizing the whole system by the Fano-type technique \cite{Fano1961PR,Barnett1988OC,huttner92},
we find the squeezed and correlated reservoirs in the ground state.
By supposing this reservoir state,
the master equation certainly guarantees the decay of the polariton system
to its original ground state in the closed case.
We also check that, when the polariton system is in the ground state,
the virtual photons in the photonic reservoir are not counted
by normal- and time-ordering the operators in polariton base.
In the input-output formalism,
we also obtain no photon detection
by supposing the squeezed and correlated reservoirs
and by normal- and time-ordering the operators.
Then, we achieve the consistency of the three approaches
(diagonalization, master equation, and input-output formalism)
even in the ultrastrong light-matter coupling regime.

This paper is organized as follows.
The Hamiltonian is shown in Sec.~\ref{sec:Hamiltonian},
and basic features of the ultrastrong coupling regime
is also discussed.
The Fano-type diagonalization of the photonic part is performed in App.~\ref{app:diag_photon}.
In Sec.~\ref{app:diag_whole}, we diagonalize the whole Hamiltonian 
and show that the reservoirs are squeezed and correlated
even in the ground state of the whole system.
The master-equation approach is discussed in Sec.~\ref{sec:master},
where we demonstrate the decay of polariton system to its original ground state.
Correlation functions of reservoir fields
are calculated in App.~\ref{app:corr_system_reservoir},
and the detailed calculation of master equation and photon detection is shown
and App.~\ref{app:ordering_master}.
The input-output approach is discussed in Sec.~\ref{sec:Langevin},
and the detailed calculation of photon detection in this approach
is shown in App.~\ref{app:ordering_input-output}.
Finally, we discuss the comparison with previous theories in Sec.~\ref{sec:discussion},
and the summary is in Sec.~\ref{sec:summary}.

\section{Hamiltonian} \label{sec:Hamiltonian}
The Hamiltonian describing cavity photons and excitations in matter is written as
\begin{equation}
\oHz = \hbar\wc\oad\oa + \hbar\wx\obd\ob + \ii\hbar\rabi(\oa+\oad)(\ob-\obd)
+ \hbar\AA(\oa+\oad)^2.
\end{equation}
Here, $\oa$ and $\ob$ are annihilation operators
of the photon and excitation, respectively,
satisfying the bosonic commutation relation
$[\oa, \oad] = [\ob, \obd] = 1$
and $[\oa,\oa] = [\ob,\ob] = [\oa,\ob] = [\oa,\obd] = 0$.
$\wc$ and $\wx$ are their eigen frequencies,
and $\rabi$ is the coupling strength.
The ultrastrong coupling means $\rabi \gtrsim \wx$.
The last term is the so-called diamagnetic term
naturally derived in the minimal coupling scheme \cite{Ciuti2005PRB},
and the coefficient is normally $\AA \ge \rabi^2/\wx$,
by which we cannot expect the quantum phase transition
\cite{Emary2004PRA,Nataf2010NC}.
If the polariton system is isolated from the environment,
as discussed in Ref.~\cite{Ciuti2005PRB},
this Hamiltonian can be diagonalized as
\begin{equation}
\oHz = \sum_{j = L, U} \hbar\omega_j \opd_j\op_j + \const
\end{equation}
Here, $\op_L$ and $\op_U$ are annihilation operators
of lower and upper polaritons, respectively.
They are represented as a combination of annihilation and creation operators
of photon and excitation:
\begin{equation} \label{eq:op=a+b+ad+bd} 
\op_j = w_j \oa + x_j \ob + y_j \oad + z_j \obd.
\end{equation}
These coefficients and eigen frequencies $\omega_j$
are determined by solving
\begin{equation} \label{eq:eigen_state_closed} 
\begin{pmatrix}
\wc+2\AA & -\ii\rabi & -2\AA & -\ii\rabi \\
\ii\rabi & \wx & -\ii\rabi & 0 \\
2\AA & -\ii\rabi & -\wc-2\AA & -\ii\rabi \\
-\ii\rabi & 0 & \ii\rabi & - \wx
\end{pmatrix}
\begin{pmatrix}
w_j \\ x_j \\ y_j \\ z_j
\end{pmatrix}
= \omega_j
\begin{pmatrix}
w_j \\ x_j \\ y_j \\ z_j
\end{pmatrix}.
\end{equation}
From this eigen value problem, we get four eigen values
$\{\omega_L, \omega_U, -\omega_L, -\omega_U\}$,
whose eigen vectors correspond to $\{\op_L, \op_U, \opd_L, \opd_U\}$,
respectively.
The coefficients are normalized for satisfying
$[\op_j, \opd_k] = \delta_{j,k}$ for $j,k = L,U$
and we also get $[\op_j, \op_k] = 0$.
Inversely, the photon and excitation operators are represented
by the polariton operators as
\begin{equation} \label{eq:eigen_closed} 
\begin{pmatrix}
\oa \\ \ob \\ \oad \\ \obd
\end{pmatrix}
= 
\begin{pmatrix}
w_L^* & w_U^* & -y_L & -y_U \\
x_L^* & x_U^* & -z_L & -z_U \\
-y_L^* & -y_U^* & w_L & w_U \\
-z_L^* & -z_U^* & x_L & x_U
\end{pmatrix}
\begin{pmatrix}
\op_L \\ \op_U \\ \opd_L \\ \opd_U
\end{pmatrix}.
\end{equation}

The photonic and excitonic reservoirs are individually
represented as ensembles of harmonic oscillators as
\begin{equation}
\oHR = \sum_m \hbar\varOmega^c_m \oalphad_m \oalpha_m
+ \sum_m \hbar\varOmega^x_m \obetad_m \obeta_m,
\end{equation}
where $\oalpha_m$ and $\obeta_m$ are annihilation operators of oscillators
in photonic and excitonic reservoirs, respectively,
and $\varOmega^{c,x}_m$ is the oscillating frequency.
The ensembles show nearly continuous spectra.
These operators satisfy $[\oalpha_m,\oalphad_n]
= [\obeta_m,\obetad_n] = \delta_{m,n}$
and $[\oalpha_m,\oalpha_n] = [\obeta_m,\obeta_n] = [\oalpha_m,\obeta_n] = [\oalpha_m,\obetad_n] = 0$.
The system-reservoir coupling is represented as
\begin{equation} \label{eq:Hint=aFc+bFx} 
\oHint = \ii\hbar(\oFd_c\oa - \oad\oF_c) + \ii\hbar(\oFd_x\ob - \obd\oF_x).
\end{equation}
Here, $\oF_c$ and $\oF_x$ are photonic and excitonic reservoir fields,
respectively, and they are expressed by the annihilation operators
$\oalpha_m$ and $\obeta_m$
and coupling strengths $\kappa_m$ and $\gamma_m$ as
\begin{subequations}
\begin{align}
\oF_c & = \sum_m \kappa_m \oalpha_m, \\
\oF_x & = \sum_m \gamma_m \obeta_m.
\end{align}
\end{subequations}
It is worth noting that Eq.~\eqref{eq:Hint=aFc+bFx} is not the result of RWA,
but this expression is naturally derived
considering the transmission and reflection of particles
between inside and outside of the cavity \cite{Dutra2000JOB}.
In other words, concerning the coupling between cavity photons
and external photonic field, they are coupled through
the electric field and also through the magnetic field.
By summing these two interaction $(\ii\hbar/2)(\oa-\oad)(\oFd_c+\oF_c)$
and $(\ii\hbar/2)(\oa+\oad)(\oFd_c-\oF_c)$,
we can derive the first term in Eq.~\eqref{eq:Hint=aFc+bFx}.
For simplicity, we also suppose the similar situation
concerning the coupling with excitonic reservoir.
Whereas the Hermitian expressions has been supposed
for describing the dissipation in some works
\cite{Scala2007JPA,Scala2007PRA,Beaudoin2011PRA,Ridolfo2012arXiv},
Eq.~\eqref{eq:Hint=aFc+bFx} can be considered as the standard expression,
because there is no ambiguity whether the system-reservoir coupling
is electric or magnetic.
Of course, when we suppose specific systems,
the expression of system-reservoir coupling is automatically determined.
In terms of the polariton operators,
the system-reservoir coupling is rewritten as
\begin{equation} \label{eq:oHint-polariton} 
\oHint = \ii\hbar(\oFd_L\op_L - \opd_L\oF_L)
+ \ii\hbar(\oFd_U\op_U - \opd_U\oF_U),
\end{equation}
where
\begin{equation} \label{eq:FLU=Fc+Fx+Fcd+Fxd} 
\oF_j = w_j \oF_c + x_j \oF_x + y_j \oFd_c + z_j \oFd_x
\end{equation}
is the external field that couples with the lower ($j=L$) or upper ($j=U$) polariton inside the cavity.

As discussed in Ref.~\cite{Ciuti2005PRB},
when we consider only the $\oHz$ system isolated from the reservoirs,
there are virtual photons and virtual excitations even in the ground state.
This is because the ground state should satisfy $\op_j\ketg = 0$,
and then the cavity mode is represented by a squeezed vacuum state
in the ground state as
\begin{subequations}
\begin{align}
\braketg{\oad\oa} & = \sum_{j=L,U} |y_j|^2, \\
\braketg{\oa\oa} & = - \sum_{j=L,U} w_j^* y_j,
\end{align}
\end{subequations}
where $\braketg{\dots}$ means an expectation value in the ground state $\ketg$.
The excitations in matter are also expressed as a squeezed vacuum state,
and the photons and excitations are correlated
in the ground state
as discussed in Ref.~\cite{Ciuti2005PRB}.
If the system-reservoir coupling is weak enough compared to
the light-matter coupling $\rabi$,
the squeezing and correlation
of cavity photon and excitation should be maintained.

\section{Diagonalization of whole system} \label{app:diag_whole}
First of all, we diagonalize the whole system $\oH = \oHz + \oHint + \oHR$
by using the Fano-type technique \cite{Fano1961PR,Barnett1988OC,huttner92}.
As shown in App.~\ref{app:diag_photon},
the photonic part consisting of cavity mode and photonic reservoir is diagonalized as
\begin{subequations} \label{eq:oHa} 
\begin{align}
\oHa & = \hbar\wc\oad\oa
+ \sum_m \hbar\varOmega^c_m\oalphad_m\oalpha_m
+ \ii\hbar(\oFd_c\oa - \oad\oF_c) \\
& = \int_0^{\infty}\dd\omega\ \hbar\omega \ooad(\omega)\ooa(\omega)
+ \text{const.}
\end{align}
\end{subequations}
Here, the partially diagonalized operator $\ooa(\omega)$
is defined in Eq.~\eqref{eq:ooa} and satisfies
\begin{equation}
\left[ \ooa(\omega), \oHa \right] = \hbar\omega \ooa(\omega)
\label{eq:[c,H]=hwc} 
\end{equation}
and
\begin{equation}
\left[ \ooa(\omega), \ooad(\omega') \right] = \delta(\omega-\omega').
\label{eq:[oc,ocd]} 
\end{equation}
Similarly, the excitonic part is diagonalized as
\begin{subequations} \label{eq:oHb} 
\begin{align}
\oHb
& = \hbar\wx\obd\ob
+ \sum_j \hbar\varOmega^x_j\obetad_j\obeta_j
+ \ii\hbar(\oFd_x\ob - \obd\oF_x) \\
& = \int_0^{\infty}\dd\omega\ \hbar\omega\oobd(\omega)\oob(\omega) + \const
\end{align}
\end{subequations}
The operator $\oob(\omega)$ is represented in Eq.~\eqref{eq:oob}.
Using these partially diagonalized operators,
the whole Hamiltonian is represented as
\begin{align}
\oH & = \int_0^{\infty}\dd\omega\ [
      \hbar\omega\ooad(\omega)\ooa(\omega) + \hbar\omega\oobd(\omega)\oob(\omega) ]
\nonumber \\ & \quad
+ \ii\hbar\rabi(\oa+\oad)(\ob-\obd)
+ \hbar\AA(\oa+\oad)^2 + \const
\end{align}
The light-matter coupling and diamagnetic terms are also 
expressed in terms of $\ooa(\omega)$ and $\oob(\omega)$
by using Eqs.~\eqref{eq:ooa2oa} and \eqref{eq:oob2ob}.
For diagonalizing the whole Hamiltonian $\oH$,
we suppose a new operator
\begin{align}
\oop(\omega) & = \int_0^{\infty}\dd\omega'\ [
  \ww(\omega,\omega') \ooa(\omega')
+ \xx(\omega,\omega') \oob(\omega')
\nonumber \\ & \quad
+ \yy(\omega,\omega') \ooad(\omega')
+ \zz(\omega,\omega') \oobd(\omega') ].
\end{align}
The coefficient functions are determined for satisfying
\begin{equation} \label{eq:[P,H]=hwP} 
[\oop(\omega), \oH] = \hbar\omega \oop(\omega)
\end{equation}
and 
\begin{equation} \label{eq:[oop,oopd]} 
[\oop(\omega), \oopd(\omega)] = \delta(\omega-\omega').
\end{equation}
Then, $\oH$ can be diagonalized as
\begin{equation}
\oH = \int_0^{\infty}\dd\omega\
      \hbar\omega\oopd(\omega)\oop(\omega) + \const
\end{equation}
From Eq.~\eqref{eq:[P,H]=hwP}, the coefficient functions are determined
by the eigen value problem
\begin{widetext}
\begin{equation}
\begin{pmatrix}
\omega'+2\AA-\omega & -\ii\rabi & -2\AA & -\ii\rabi \\
\ii\rabi & \omega'-\omega & -\ii\rabi & 0 \\
2\AA & -\ii\rabi & -\omega'-2\AA-\omega & -\ii\rabi \\
-\ii\rabi & 0 & \ii\rabi & - \omega'-\omega
\end{pmatrix}
\begin{pmatrix}
\uc(\omega')\ww(\omega,\omega') \\
\ux(\omega')\xx(\omega,\omega') \\
\uc(\omega')^*\yy(\omega,\omega') \\
\ux(\omega')^*\zz(\omega,\omega')
\end{pmatrix}
= 0,
\end{equation}
where coefficients $u_{c,x}(\omega)$ is represented in Eq.~\eqref{eq:uc}.
This eigen value problem is equivalent to Eq.~\eqref{eq:eigen_closed}
by replacing $\omega_c$ and $\omega_x$ with $\omega'$,
and the eigen frequencies $\omega_{L,U}(\omega')$ must be equal to $\omega$.
Then, operator $\oop(\omega)$ is represented as
\begin{align}
\oop(\omega)
& = \sum_{j=L,U} \left\{
  \frac{w_j(\winv_j)}{\uc(\winv_j)} \ooa(\winv_j)
+ \frac{x_j(\winv_j)}{\ux(\winv_j)} \oob(\winv_j)
+ \frac{y_j(\winv_j)}{\uc(\winv_j)^*} \ooad(\winv_j)
+ \frac{z_j(\winv_j)}{\ux(\winv_j)^*} \oobd(\winv_j)
\right\}.
\end{align}
\end{widetext}
Here, $\winv_j$ is the frequency satisfying $\omega = \omega_j(\winv_j)$,
and $w_j(\omega)$, $x_j(\omega)$, $y_j(\omega)$, and $z_j(\omega)$ are
the coefficients when we solve Eq.~\eqref{eq:eigen_closed}
by replacing $\omega_c$ and $\omega_x$ with $\omega$.
They are normalized for satisfying Eq.~\eqref{eq:[oop,oopd]}.
Inversely, we can rewrite $\ooa(\omega)$ and $\oob(\omega)$ as
\begin{subequations}
\begin{align}
\ooa(\omega) & = \uc(\omega) \sum_{j=L,U} \left[
  w_j(\omega)^* \oop(\omega_j(\omega)) - y_j(\omega) \oopd(\omega_j(\omega))
\right], \\
\oob(\omega) & = \ux(\omega) \sum_{j=L,U} \left[
  x_j(\omega)^* \oop(\omega_j(\omega)) - z_j(\omega) \oopd(\omega_j(\omega))
\right].
\end{align}
\end{subequations}
Then, the original photon, excitation, and reservoir operators
are also expressed in terms of $\oop(\omega)$
by using Eqs.~\eqref{eq:ooa2oa} and \eqref{eq:oob2ob}.

The ground state $\ketgg$ of the whole system $\oH$ is determined for
satisfying $\oop(\omega)\ketgg = 0$ for $0 < \omega < \infty$.
Then, when we apply the photonic free field
\begin{equation}
\oFz_c(\tau) = \ee^{\ii\oHR t} \oFz_c \ee^{-\ii\oHR t}
\end{equation}
onto the ground state,
it is represented as
\begin{align}
\oFz_c(\tau)\ketgg
& = - \int_0^{\infty}\dd\omega\int_0^{\infty}\dd\omega'\
  \ee^{-\ii\omega\tau}\kappa(\omega)v_c(\omega',\omega)^* u_c(\omega')
\nonumber \\ & \quad \times
  \sum_{j=L,U} y_j(\omega')\oopd(\omega_j(\omega')) \ketgg,
\end{align}
where coefficient $v_c(\omega',\omega)$ is expressed in Eq.~\eqref{eq:vc}.
The phase-independent correlation function is written as
\begin{align}
& \bragg \oFzd_c\oFz_c(\tau) \ketgg
\nonumber \\ &
= \int_0^{\infty}\dd\omega\int_0^{\infty}\dd\omega'\int_0^{\infty}\dd\omega''\
  \ee^{-\ii\omega'\tau} \kappa(\omega)^*\kappa(\omega')
\nonumber \\ & \quad \times
  v_c(\omega'',\omega)v_c(\omega'',\omega')^*
  |u_c(\omega'')|^2 \sum_{j=L,U} |y_j(\omega'')|^2.
\end{align}
Here, the coefficients $\uc(\omega)$ and $\vc(\omega,\omega')$ are singular
at $\omega = \wc$ as seen in Eqs.~\eqref{eq:uc} and \eqref{eq:vc}.
Since the coefficients are normalized as
\begin{align}
\int_0^{\infty}\dd\omega\ |\uc(\omega)|^2 & = 1, \\
\int_0^{\infty}\dd\omega\ \vc(\omega,\omega')\vc(\omega,\omega'') & = \delta(\omega'-\omega''),
\end{align}
if the dissipation of photons is weak enough
compared to the characteristic frequencies of $\oHz$ system
($\rabi$, $\wc$, and $\wx$),
the correlation function is approximately represented in the ground state as
\begin{align}
& \bragg \oFzd_c\oFz_c(\tau) \ketgg
\nonumber \\ &
\simeq \int_0^{\infty}\dd\omega\
  \ee^{-\ii\omega\tau} |\kappa(\omega)|^2
  \sum_{j=L,U} |y_j(\wc)|^2
= G_c(\tau) \braketg{\oad\oa}.
\end{align}
In the same manner, the phase-sensitive correlation is expressed
in the ground state as
\begin{align}
& \bragg \oFz_c\oFz_c(\tau) \ketgg
\nonumber \\ &
= - \int_0^{\infty}\dd\omega\int_0^{\infty}\dd\omega'\int_0^{\infty}\dd\omega''\
  \ee^{-\ii\omega'\tau} \kappa(\omega)\kappa(\omega')
\nonumber \\ & \quad \times
  v_c(\omega'',\omega)^*v_c(\omega'',\omega')^*
  u_c(\omega'')^2 \sum_{j=L,U} w_j(\omega'')^*y_j(\omega'') \nonumber \\
& \simeq - \int_0^{\infty}\dd\omega\
  \ee^{-\ii\omega\tau} |\kappa(\omega)|^2
  \sum_{j=L,U} w_j(\wc)^* y_j(\wc)
\nonumber \\ &
= G_c(\tau) \braketg{\oa\oa}.
\end{align}
Therefore, even in the ground state,
the photonic reservoir field $\oF_c$ is also squeezed
by the same order as the cavity mode.
In the same manner,
the correlation of the two reservoir fields
are also the same as internal ones
if the dissipation is weak enough.
In other words, the internal modes $\oa$ and $\ob$ are balanced with $\oF_c$ and $\oF_x$,
respectively, in the ground state of the whole system.

When we initially suppose a squeezed cavity mode
and a big reservoir in the vacuum state,
of course the reservoir does not become equally squeezed
but it almost remains in the vacuum state
after switching on the coupling between them.
This is a non-equilibrium problem.
However, when we consider the equilibrium of the cavity mode
and the reservoir, if the ground state of the cavity mode is squeezed,
the reservoir field is also squeezed in the ground state of the whole system.
Therefore, when we consider the dissipation of the ultrastrong light-matter coupling system
to its original ground state,
we should suppose the squeezed and correlated reservoirs
as will be discussed in the following two sections and also in Sec.~\ref{sec:discussion}.

\section{Master-equation approach} \label{sec:master}
In this section, we derive a master equation
for describing the dissipation of polariton system $\oHz$
by considering the coupling with reservoirs.
Obeying the standard derivation of master equations \cite{Breuer2006,Scala2007PRA,Scala2007JPA,Fleming2010JPA,Nakatani2010JPSJ,Beaudoin2011PRA},
from the expression \eqref{eq:oHint-polariton} of system-reservoir coupling $\oHint$,
the master equation for reduced density operator $\orho(t)$ describing $\oHz$ system
is derived under the Born approximation as
\begin{equation} \label{eq:master_Born} 
\ddt{}\orho(t) = \oL[\orho],
\end{equation}
where $\oL = \oL_0 + \oLdiss$ and
\begin{equation}
\oL_0[\orho] = \frac{1}{\ii\hbar}[\oHz, \orho(t)],
\end{equation}
\begin{align} \label{eq:oLdiss_Born} 
\oLdiss[\orho] 
& = \sum_{j,k=L,U} \left\{
  \left[\oDL_{jk}[\orho], \opd_j\right]
+ \left[\op_j, \oDR_{jk}[\orho] \right]
\right. \nonumber \\ & \quad
+ \left[\oCL_{jk}[\orho], \op_j\right]
+ \left[\opd_j, \oCR_{jk}[\orho] \right]
+ \left[\opd_j, \oBL_{jk}[\orho]\right]
\nonumber \\ & \quad \left.
+ \left[\oBR_{jk}[\orho], \opd_j \right]
+ \left[\op_j, \oAL_{jk}[\orho]\right]
+ \left[\oAR_{jk}[\orho], \op_j \right]
\right\},
\end{align}
\begin{subequations}
\begin{align}
\oDL_{jk}[\orho]
& = \int_{t_0}^{t}\dd t'\ \oUz(t-t')[\op_k \orho(t')]
  \braketin{\oFz_{j}(t)\oFzd_{k}(t')} \\
\oDR_{jk}[\orho]
& = \int_{t_0}^{t}\dd t'\ \oUz(t-t')[\orho(t') \opd_k]
  \braketin{\oFz_{k}(t')\oFzd_{j}(t)} \\
\oCL_{jk}[\orho]
& = \int_{t_0}^{t}\dd t'\ \oUz(t-t')[\opd_k\orho(t')]
  \braketin{\oFzd_{j}(t)\oFz_{k}(t')} \\
\oCR_{jk}[\orho]
& = \int_{t_0}^{t}\dd t'\ \oUz(t-t')[\orho(t') \op_k]
  \braketin{\oFzd_{k}(t')\oFz_{j}(t)} \\
\oBL_{jk}[\orho]
& = \int_{t_0}^{t}\dd t'\ \oUz(t-t')[\opd_k\orho(t')]
  \braketin{\oFz_{j}(t)\oFz_{k}(t')} \\
\oBR_{jk}[\orho]
& = \int_{t_0}^{t}\dd t'\ \oUz(t-t')[\orho(t') \opd_k]
  \braketin{\oFz_{k}(t')\oFz_{j}(t)} \\
\oAL_{jk}[\orho]
& = \int_{t_0}^{t}\dd t'\ \oUz(t-t')[\op_k \orho(t')]
  \braketin{\oFzd_{j}(t)\oFzd_{k}(t')} \\
\oAR_{jk}[\orho]
& = \int_{t_0}^{t}\dd t'\ \oUz(t-t')[\orho(t') \op_k]
  \braketin{\oFzd_{k}(t')\oFzd_{j}(t)}
\end{align}
\end{subequations}
Here, $t_0 \rightarrow -\infty$ is the switch-on time of system-reservoir coupling.
$\oUz(\tau)[\oO]$ is the propagator in $\oHz$ system
for arbitrary operator $\oO$ as
\begin{equation}
\oUz(\tau)[\oO] = \ee^{-\ii\oHz\tau/\hbar} \oO \ee^{\ii\oHz\tau/\hbar},
\end{equation}
and $\oFz_{L,U}$ is the reservoir field in the interaction picture
(free field):
\begin{equation} \label{eq:oFzLU} 
\oFz_{L,U}(t) = \ee^{\ii\oHR t} \oF_{L,U} \ee^{-\ii\oHR t}.
\end{equation}
These free fields are in the polariton base, and it is represented
by the ones $\oFz_{c,x}$ in the excitation-photon base
as in Eq.~\eqref{eq:FLU=Fc+Fx+Fcd+Fxd}.
They satisfy ($\mu,\nu = c, x$)
\begin{subequations} \label{eq:[oFz,oFzd]} 
\begin{align}
[\oFz_{\mu}(t), \oFzd_{\nu}(t')] & = \delta_{\mu,\nu} G_{\mu}(t-t'), \\
[\oFz_{\mu}(t), \oFz_{\nu}(t')] & = 0.
\end{align}
\end{subequations}
where $\Gc(\tau)$ and $\Gx(\tau)$ are memory kernels
due to the coupling with reservoirs and are expressed as
\begin{subequations}
\begin{align}
\Gc(\tau) & = \sum_m |\kappa_m|^2 \ee^{-\ii\varOmega_m^c\tau}, \\
\Gx(\tau) & = \sum_m |\gamma_m|^2 \ee^{-\ii\varOmega_m^x\tau}.
\end{align}
\end{subequations}
From the master equation \eqref{eq:master_Born},
the dynamics in $\oHz$ system are determined
by supposing correlation functions of the free fields,
which are considered as an input from the reservoirs to the $\oHz$ system.

\begin{figure}
\begin{center}
\includegraphics[width=.5\textwidth]{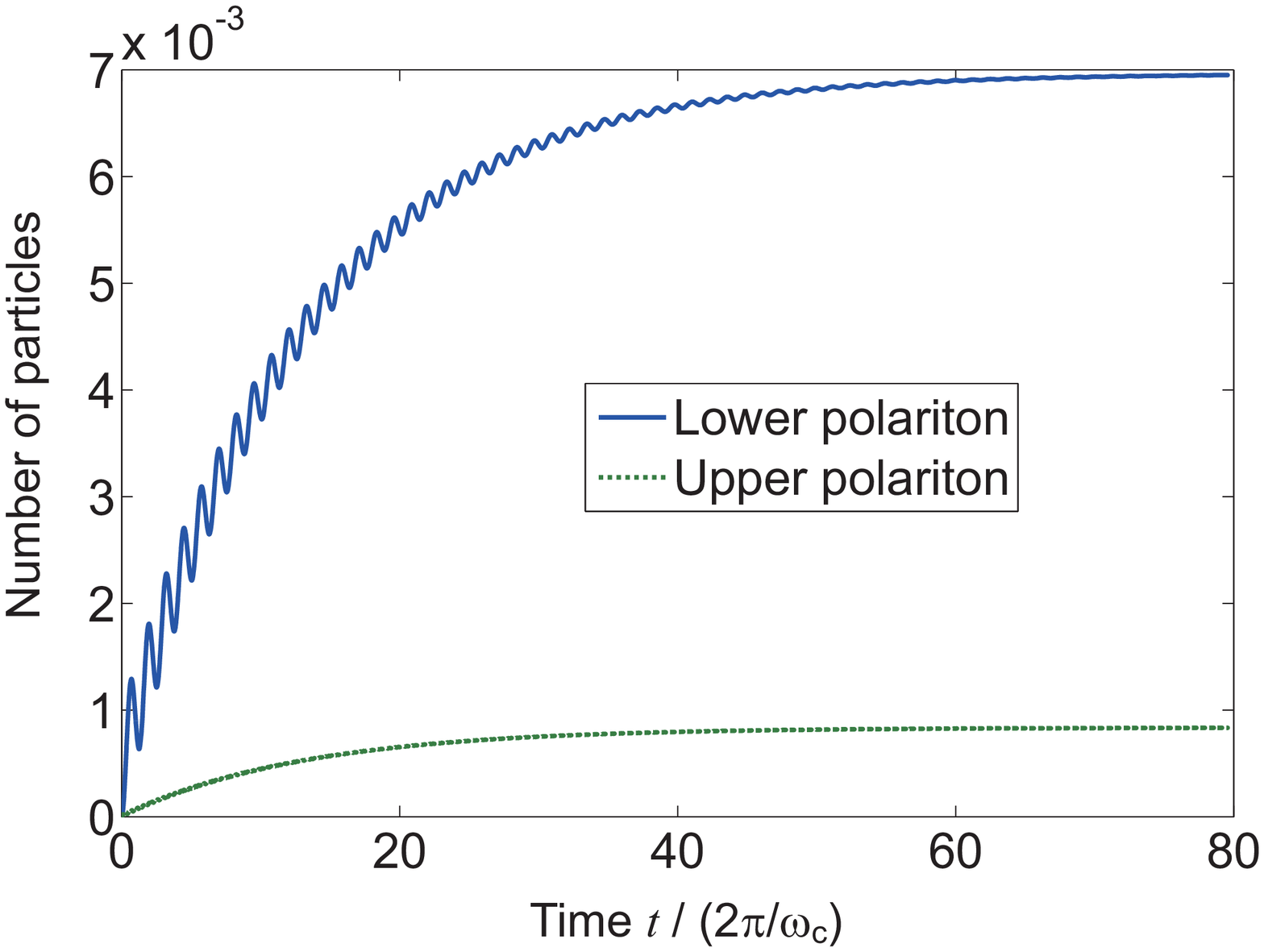}
\end{center}
\caption{Numbers of lower and upper polaritons are plotted as a function
of time $t$. The initial state is given as $\orho(0) = \ketg\brag$,
and the reservoirs are supposed to be in the vacuum state.
The time-development is calculated by master equation
\eqref{eq:master_Born}, correlation \eqref{eq:<FzFz>-vacuum},
and memory kernel \eqref{eq:kernel-flat}.
Parameters: $\wx = \wc$, $\rabi = \wc$, $\AA = \rabi^2/\wx$,
$\varGamma_c = \varGamma_x = 10^{-2}\wc$, and $\wco_c = \wco_x = 10^3\wc$.}
\label{fig:1}
\end{figure}
First of all, we suppose that the photonic and excitonic reservoirs
are in the vacuum state and the correlation functions are given as
($\mu,\nu = c, x$)
\begin{subequations} \label{eq:<FzFz>-vacuum} 
\begin{align}
\braketin{\oFz_{\mu}(t)\oFzd_{\nu}(t')} & = \delta_{\mu,\nu} G_{\mu}(t-t'), \\
\braketin{\oFzd_{\mu}(t)\oFz_{\nu}(t')}
& = \braketin{\oFz_{\mu}(t)\oFz_{\nu}(t')} = 0.
\end{align}
\end{subequations}
In Fig.~\ref{fig:1},
supposing the ground state $\orho(0) = \ketg\brag$ at the initial time $t = 0$,
we plot the development of numbers of lower and upper polaritons
calculated by the master equation \eqref{eq:master_Born}
and the correlation \eqref{eq:<FzFz>-vacuum}.
The memory kernels are simply supposed as
\begin{subequations} \label{eq:kernel-flat} 
\begin{align}
\Gc(\tau) & = \int_0^{\wco_c}\dd\varOmega\ \frac{\varGamma_c}{2\pi} \ee^{-\ii\varOmega\tau}, \\
\Gx(\tau) & = \int_0^{\wco_x}\dd\varOmega\ \frac{\varGamma_x}{2\pi} \ee^{-\ii\varOmega\tau},
\end{align}
\end{subequations}
where the cut-off frequency $\wco_{c,x}$ governs the memory time of the reservoirs as
$\sim 1/\wco_{c,x}$.
In a similar way as in Ref.~\cite{DeLiberato2009PRA},
the density operator $\orho(t)$ is moved outside the time integral
in the master equation \eqref{eq:master_Born}.
This treatment is valid if the memory time $1/\wco_{c,x}$ is short enough
compared to the specific oscillation periods ($1/\wc$, $1/\wx$, and $1/\rabi$)
of $\oHz$ system.

As seen in Fig.~\ref{fig:1},
the polaritons are excited by the vacuum reservoirs (at zero temperature).
The periods of the oscillation are approximately $\pi/\omega_{L,U}$
($\omega_L/\wc = 0.414$ and $\omega_U/\wc = 2.414$),
whereas they are slightly modified by the Lamb shifts.
After a long time compared to $1/\varGamma_c = 1/\varGamma_x = (100/2\pi)\times(2\pi/\wc)$,
the numbers of polaritons reach to certain values,
which depend on the system-reservoir coupling strengths $\varGamma_{c,x}$.

The polariton system is excited by the vacuum reservoirs,
because it is excited when the virtual photons in the ground state
escape to the reservoirs.
In other words, the ground state of polariton system is modified
by the coupling with vacuum reservoirs.
This result can also be understood by Eq.~\eqref{eq:oLdiss_Born}.
In order to guarantee the decay of the $\oHz$ system
to its original ground state $\ketg$,
the photonic and excitonic reservoirs should not be in the vacuum state
in the excitation-photon base (in terms of $\oFz_{c,x}$),
but the free fields $\oFz_{L,U}$ in polariton base should be
in the vacuum state.
In Refs.~\cite{Beaudoin2011PRA,Ridolfo2012arXiv},
owing to the RWA on system-reservoir coupling,
the decay to the ground state $\ketg$ is guaranteed
by simply considering the vacuum reservoirs in the excitation-photon base.
However, in the present paper,
we do not use the RWA to maintain the information of quantum fluctuation
of the reservoirs.
Instead, we suppose
that the reservoirs are in the vacuum state in polariton base
(squeezed and correlated in excitation-photon base).

Let's derive the correlation of reservoir free fields $\oFz_{L,U}$
that guarantees the decay to the ground state $\ketg$ of $\oHz$ system
and is simultaneously appropriate to the analysis of Fano-type diagonalization
discussed in Sec.~\ref{app:diag_whole}.
We assume that the $\oHz$ system is in the ground
state as $\orho = \ketg\brag$. 
Under this assumption, let's inversely consider
how the reservoirs are modified by coupling with $\oHz$ system.
As discussed in Ref.~\cite{Carmichael1987JOSAB}
and in App.~\ref{app:corr_system_reservoir} of this paper,
we can derive the correlation of free fields $\oFz_{L,U}$ (on output side)
from the density operator $\orho(t)$ of $\oHz$ system.
The equations of motion (Langevin equations)
of cavity photons and excitations are derived as
\begin{subequations} \label{eq:motion-a-b} 
\begin{align}
\ddt{}\oa(t) & = \frac{1}{\ii\hbar}[\oa,\oHz](t)
  - \int_{t_0}^t\dd t'\ \Gc(t-t') \oa(t') - \oFz_c(t),\\
\ddt{}\ob(t) & = \frac{1}{\ii\hbar}[\ob,\oHz](t)
  - \int_{t_0}^t\dd t'\ \Gx(t-t') \ob(t') - \oFz_x(t).
\end{align}
\end{subequations}
In the standard theory of quantum optics \cite{gardiner04,walls08},
the memory kernels $G_{c,x}(\tau)$
are approximately described by the Dirac's delta function
by elongating the frequency range of reservoirs to $-\infty$ and $\infty$.
Since the escaped photons do not reenter into a cavity,
we can consider that the photonic reservoir has a quite small coherence time,
and this approximation seems valid in most cases.
However, in the ultrastrong coupling regime,
when we do not use the RWA on system-reservoir coupling,
we must keep the reservoir frequencies positive \cite{Ciuti2006PRA},
and the Langevin and master equations are written in the time nonlocal forms
in general.
Although in the case of time-local equations
we usually use the standard input-output relation \cite{gardiner04,walls08},
we calculate the correlation between the free fields and internal ones
by the formalism of Ref.~\cite{Carmichael1987JOSAB}.
First, we define the propagator $\oU(\tau)[\cdots]$ satisfying
\begin{equation}
\orho(t+\tau) = \oU(\tau)[\orho(t)],
\end{equation}
\begin{equation}
\frac{\partial}{\partial\tau} \oU(\tau) = \oL[\oU(\tau)],
\end{equation}
and the quantum regression theorem
\cite{Lax1963PR,Lax1967PR,Carmichael1987JOSAB,gardiner04,Breuer2006,walls08}
is written for $\tau > 0$ as
\begin{subequations} \label{eq:QRT} 
\begin{align}
\braket{\oO_1(t+\tau)\oO_2(t)} & = \Tr\{\oO_1\oU(\tau)[\oO_2\orho(t)]\}, \\
\braket{\oO_1(t)\oO_2(t+\tau)} & = \Tr\{\oO_2\oU(\tau)[\orho(t)\oO_1]\}.
\end{align}
\end{subequations}
As discussed in detail in App.~\ref{app:corr_system_reservoir},
by using this and Eq.~\eqref{eq:motion-a-b},
the correlation between $\oFz_{\mu}(t)$ and arbitrary operator $\oS(t)$ in $\oHz$ system is derived as ($\mu=c,x$)
\begin{subequations} \label{eq:<SFz(t)>} 
\begin{align}
\braket{\oS(t)\oFz_{\mu}(t+\tau)}
& = - \int_{t_0}^{t}\dd t'\ G_{\mu}(t+\tau-t')\braket{\oS(t)\os_{\mu}(t')},
\label{eq:<S(t)Fz(t+tau)>} \\ 
\braket{\oFz_{\mu}(t+\tau)\oS(t)}
& = - \int_{t_0}^{t}\dd t'\ G_{\mu}(t+\tau-t')\braket{\os_{\mu}(t')\oS(t)},
\label{eq:<Fz(t+tau)S(t)>} 
\end{align}
\end{subequations}
\begin{subequations} \label{eq:<S(t)Fz>} 
\begin{multline}
\braket{\oS(t+\tau)\oFz_{\mu}(t)}
= - \int_{t_0}^{t}\dd t'\ G_{\mu}(t-t')\braket{\oS(t+\tau)\os_{\mu}(t')} \\
    - \Tr\{\oS\oU(\tau)[\os_{\mu}\oLdiss[\orho(t)] - \oLdiss[\os_{\mu}\orho(t)]]\},
\end{multline}
\begin{multline}
\braket{\oFz_{\mu}(t)\oS(t+\tau)}
= - \int_{t_0}^{t}\dd t'\ G_{\mu}(t-t')\braket{\os_{\mu}(t')\oS(t+\tau)} \\
    - \Tr\{\oS\oU(\tau)[\oLdiss[\orho(t)]\os_{\mu} - \oLdiss[\orho(t)\os_{\mu}]]\},
\end{multline}
\end{subequations}
where
\begin{equation}
\os_{\mu} = \begin{cases} \oa & \text{for $\mu = c$} \\ \ob & \text{for $\mu = x$} \end{cases}
\end{equation}
Whereas Eqs.~\eqref{eq:<SFz(t)>} are zero in the limit of time-local case
$G_{\mu}(\tau) \propto \delta(\tau)$ \cite{Carmichael1987JOSAB},
they are in general non-zero in the present nonlocal situation.
Of course, if $\tau$ is large enough compared to the memory time of the reservoirs,
Eqs.~\eqref{eq:<SFz(t)>} is negligible
compared to Eqs.~\eqref{eq:<S(t)Fz>}.
Furthermore, the self-correlation of free fields $\oFz_{c,x}$
is obtained in a steady state (the ground state in the present case) as
\begin{subequations} \label{eq:<Fz(t)Fz>} 
\begin{align}
\braket{\oFzd_{\mu}(\tau)\oFz_{\nu}} & = G_{\mu}^*(\tau) \braketg{\osd_{\mu}\os_{\nu}} \\
\braket{\oFz_{\mu}(\tau)\oFzd_{\nu}} & = G_{\mu}(\tau) \braketg{\os_{\mu}\osd_{\nu}} \\
\braket{\oFz_{\mu}(\tau)\oFz_{\nu}} & = \begin{cases}
G_{\mu}(|\tau|) \braketg{\os_{\mu}\os_{\nu}} & \text{for $\tau > 0$} \\
G_{\nu}(|\tau|) \braketg{\os_{\mu}\os_{\nu}} & \text{for $\tau < 0$}
\end{cases}
\end{align}
\end{subequations}
where $\braketg{\cdots}$ means an expectation value in the steady state
(ground state).
These correlation certainly satisfies Eqs.~\eqref{eq:[oFz,oFzd]},
and is equivalent to the ones derived in Sec.~\ref{app:diag_whole}.

In the sense of perturbation theory, they are the correlation on output side,
i.e., the modification of the reservoirs due to the coupling with $\oHz$ system.
The free-field correlation
appearing in the master equation \eqref{eq:oLdiss_Born}
is the one on the input side (effect from reservoirs to $\oHz$ system).
Here, under the equilibrium between $\oHz$ and reservoirs,
the correlation of $\oFz_{c,x}$ should be equivalent on both input and output side.
Let's substitute Eqs.~\eqref{eq:<Fz(t)Fz>}
to the master equation \eqref{eq:oLdiss_Born}.
The free-field correlation in the polariton base
can be derived by using Eq.~\eqref{eq:FLU=Fc+Fx+Fcd+Fxd}.
From Eqs.~\eqref{eq:<Fz(t)Fz>},
we can easily get for $j,k = L,U$
\begin{equation}
\braketin{\oFz_j(\tau>0)\oFz_k}
= \braketin{\oFzd_j(\tau>0)\oFz_k}
= 0,
\end{equation}
and then the master equation is reduced to
\begin{align}
\ddt{}\orho(t) & = 
\oL_0[\orho]
+ \sum_{j,k=L,U} \left\{
  \left[\oDL_{jk}[\orho]+\oBR_{jk}[\orho], \opd_j\right]
\right. \nonumber \\ & \quad \left.
+ \left[\op_j, \oDR_{jk}[\orho]+\oAL_{jk}[\orho] \right]
\right\}. \label{eq:master_Born_diss} 
\end{align}
The steady state obtained from this equation
is certainly the ground state of closed case $\orhoss = \ketg\brag$,
then the decay of polariton system to its original ground state $\ketg$
is guaranteed by supposing the squeezed and correlated reservoirs in Eqs.~\eqref{eq:<Fz(t)Fz>}.

\begin{figure}
\begin{center}
\includegraphics[width=.5\textwidth]{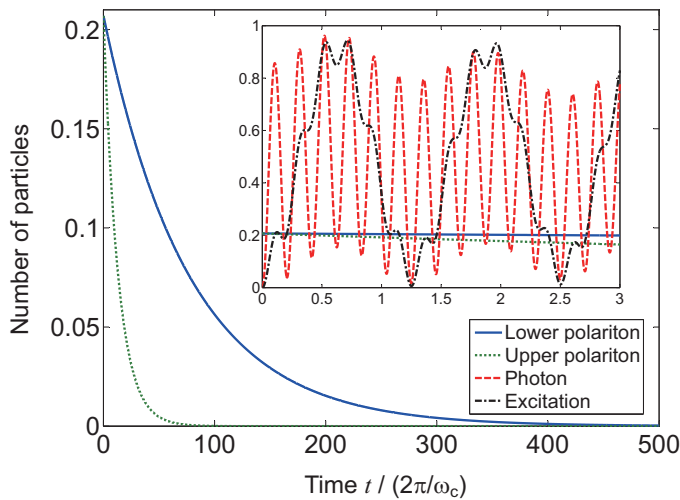}
\end{center}
\caption{Starting from the vacuum state of photons and excitations,
the numbers of polaritons are calculated as a function of time
by master equation \eqref{eq:master_Born_diss},
which are derived by the correlated and squeezed reservoirs
as in Eqs.~\eqref{eq:<Fz(t)Fz>}.
In the inset, the numbers of photons and excitations are also plotted
in the early stage.
Parameters: $\wx = \wc$, $\rabi = \wc$, $\AA = \rabi^2/\wx$,
$|\kappa|^2 = |\gamma|^2 = 10^{-2}\wc/2\pi$, and $\wco_c = \wco_x = 10^3\wc$.}
\label{fig:2}
\end{figure}
By using this master equation \eqref{eq:master_Born_diss},
we have calculated the dynamics of $\oHz$ system.
In Fig.~\ref{fig:2}, 
supposing the vacuum state (no photon and no excitation) at the initial time $t = 0$,
the numbers of lower and upper polaritons are plotted as a function of time.
In the numerical simulation,
the density operator $\orho(t)$ is moved outside the time integral,
and the memory kernels are also given in Eq.~\eqref{eq:kernel-flat}.
While there are non-zero polaritons at the initial time,
the numbers of polaritons decrease and finally go to zero,
i.e., the $\oHz$ system decays to its ground state $\ketg$.
In the inset of Fig.~\ref{fig:2},
we also plot the numbers of photons and excitations in the early stage.
Whereas both of them are zero at the initial time $t = 0$,
they are oscillated with two periods $\pi/\omega_L$ and $\pi/\omega_U$
(slightly modified by the Lamb shifts),
but finally they reach to $\braketg{\oad\oa} = \braketg{\obd\ob}
= 0.207$ after a long time (not shown in the figure).

Under the Born approximation, the total density operator $\orhotot$
is approximately represented by the product of the density operator $\orho$
of $\oHz$ system and the one $\orhoR$ of reservoirs as $\orhotot = \orho \otimes \orhoR$.
If the system-reservoir coupling is weak enough for the Born approximation,
in the ground state $\ketgg$ of the whole system,
the state of $\oHz$ system is approximately equivalent to the ground state $\ketg$
of the closed case.
On the other hand, the free-field correlation \eqref{eq:<Fz(t)Fz>} approximately
reflects the reservoir state that is obtained
by tracing over the $\oHz$ variables on the ground state $\ketgg$
as $\orhoR \simeq \Tr_S\{\ketgg\bragg\}$,
which was verified in Sec.~\ref{app:diag_whole}.
This reservoir state is not the ground state of $\oHR$,
but it certainly guarantees the decay of $\oHz$ system
to its original ground state $\ketg$ as seen in Fig.~\ref{fig:2}.
If we suppose the ground state of $\oHR$ system,
in which photonic and excitonic reservoirs are in vacuum (at zero temperature),
the $\oHz$ system does not decay to its ground state $\ketg$
as seen in Fig.~\ref{fig:1}.
However, the obtained steady state is approximately equivalent to the ground state,
if the system-reservoir coupling is weak enough.
Next, let's calculate the output from the cavity
in the formalism of master equation.
If the $\oHz$ system is in the ground state,
we cannot detect anything outside the cavity.

\begin{figure}
\begin{center}
\includegraphics[width=.3\textwidth]{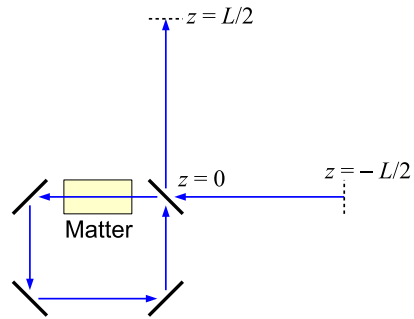}
\end{center}
\caption{Sketch of ring-cavity system.
Inside the cavity, photons interact with matter,
but back scattering of photons does not occur.
The external field is defined in the one dimensional system with length $L$,
and the field is continuously connected at the boundaries $z = \pm L/2$.}
\label{fig:ring_cavity}
\end{figure}
As seen in Fig.~\ref{fig:ring_cavity}, we consider a ring-shape cavity
embedding a matter interacting with photons inside the cavity
as discussed in Ref.~\cite{Carmichael1987JOSAB}.
We assume that back scattering
of photons does not occur during the light-matter interaction, and the clockwise and counter-clockwise fields are separated.
Concerning the external field, we consider a one dimensional system with length $L$, and the field is continuous at the two ends $z=\pm L/2$. Whereas the external photonic modes are characterized by wavenumber $k_j = 2\pi j / L$ for $j = 0,\ \pm 1,\ \pm2,\ \ldots$, the forward field $j > 0$ and backward $j < 0$ fields can be separated into independent subspaces.
Here, we focus on the forward field $j > 0$,
and its frequency is represented as $\varOmega^c_j = c k_j$,
where $c$ is the speed of light.
The density of states (DOS) is $\DOSc = L/2\pi c$.
This forward field couples with the counter-clockwise intracavity photons.
We define the propagating field in forward direction at position $z$
in the external system as
\begin{equation}
\oFfwd_c(z,t) = \sum_m \kappa_m \oalpha_m(t) \ee^{\ii k_m z}.
\end{equation}
From the equation of motion of $\oalpha_m(t)$, this field is rewritten as
\begin{align}
\oFfwd_c(z,t)
& = \sum_m \kappa_m \oalpha_m(t_0) \ee^{-\ii\varOmega_m(t - z/c - t_0)}
\nonumber \\ & \quad
+ \int_{t_0}^{t}\dd t'\ \Gc(t-t'-z/c) \oa(t').
\end{align}
As discussed in Ref.~\cite{Carmichael1987JOSAB},
by choosing a position of observation $z_0 > 0$,
we define the output field as
\begin{align}
\oFout_c(t)
& = \oFfwd_c(z_0,t+z_0/c)
\nonumber \\ &
= \oFz_c(t) + \int_{t_0}^{t+z_0/c}\dd t'\ \Gc(t-t') \oa(t').
\label{eq:def-iFout} 
\end{align}
Here, $\oFz_c(t)$ is the free field appearing in the Langevin equation
\eqref{eq:motion-a-b} and also in the master equation \eqref{eq:master_Born}.
The second term is the contribution from the cavity.
Whereas this term includes the information of cavity photons at time $t + z_0/c > t$,
the causality is not violated,
because the output field $\oFout_c(t)$ is actually
the propagating field at position $z_0$ and at time $t+z_0/c$.
In the time-local limit $\Gc(\tau) = \varGamma_c\delta(\tau)$,
Eq.~\eqref{eq:def-iFout} is correctly reduced to the well-known input-output relation \cite{gardiner04,walls08,Carmichael1987JOSAB}.
Further, in the limit of $z_0 \rightarrow \infty$ and $t_0 \rightarrow -\infty$,
Eq.~\eqref{eq:def-iFout} is reduced to the input-output relation \eqref{eq:input-output-F}
in the time-nonlocal case, 
which will be derived in Sec.~\ref{sec:Langevin}.

When we evaluate the output measured by photon detectors,
the expectation values should be normal-ordered and time-ordered
(expressed as $\braket{:\ldots:}$)
in terms of polariton operators (not of photon and excitation).
The correlation between cavity photons and the free field of photonic reservoir
can be evaluated by Eqs.~\eqref{eq:<SFz(t)>} and \eqref{eq:<S(t)Fz>},
and the self-correlation of the free field is also given
by Eqs.~\eqref{eq:<Fz(t)Fz>}.
The detail of the calculation is shown in App.~\ref{app:ordering_master}.
When we suppose that the $\oHz$ system is in the ground state $\ketg$
by considering the reservoir correlation \eqref{eq:<Fz(t)Fz>}
in the master equation,
we have numerically checked that the emission spectrum $\braket{:\oFout_c(\omega)\oFoutd_c:}$
and phase-sensitive correlation $\braket{:\oFout_c(\omega)\oFout_c:}$
are approximately zero.
The deviation is due to the approximation
that we used in the numerical calculation
(density operator $\orho(t)$ is moved outside the integral),
and it is not caused by the supposed correlation, Eqs.~\eqref{eq:<Fz(t)Fz>}.
On the other hand, if we suppose the vacuum photonic and excitonic reservoirs,
we cannot find a policy which guarantees no photon detection,
although the vacuum output is obtained for vacuum input
in the input-output formalism \cite{Ciuti2006PRA}.
This is because of the perturbation treatment in the formalism of master equation
as we will discuss in Sec.~\ref{sec:discussion}.

In this way, when we suppose the squeezed and correlated reservoir fields
as in Eqs.~\eqref{eq:<Fz(t)Fz>}, we have successfully obtained the natural result:
the $\oHz$ system decays to its ground state $\ketg$,
and the photon emission is not detectable if the system is in the ground state.
Furthermore, it is also consistent to the analysis of Fano-type diagonalization
(there are virtual photons and excitations in the reservoirs, and photonic and excitonic reservoirs are correlated with each other and also squeezed).

\section{Input-output approach} \label{sec:Langevin}
Another approach for describing the dissipation and emission
of photons is the formalism of Langevin equations with input-output relation.
As discussed in Ref.~\cite{Ciuti2006PRA},
the Langevin equations of cavity photons and excitations
are derived in frequency domain as
\begin{equation} \label{eq:Langevin_ex_ph} 
[\mM(\omega)-\omega\munit]
\begin{bmatrix}
\oa(\omega) \\ \ob(\omega) \\ \oa(-\omega)^{\dagger} \\ \ob(-\omega)^{\dagger}
\end{bmatrix}
= \ii
\begin{bmatrix}
\oFin_c(\omega) \\ \oFin_x(\omega) \\ \oFin_c(-\omega)^{\dagger} \\ \oFin_x(-\omega)^{\dagger}
\end{bmatrix}.
\end{equation}
Here, the coefficient matrix is written as
\begin{widetext}
\begin{equation} \label{eq:mM} 
\mM(\omega)
= \begin{bmatrix}
\wc+2\AA-\ii \Gc(\omega)_+ & \ii\rabi & 2\AA & -\ii\rabi \\
-\ii\rabi & \wx-\ii\Gx(\omega)_+ & -\ii\rabi & 0 \\
-2\AA & -\ii\rabi & -\wc-2\AA-\ii \Gc(-\omega)_+^* & \ii\rabi \\
-\ii\rabi & 0 & -\ii\rabi & -\wx-\ii\Gx(-\omega)_+^*
\end{bmatrix},
\end{equation}
\end{widetext}
and the memory kernels $G_{c,x}(\tau)$
are Fourier-transformed for positive time as
\begin{equation}
G_{c,x}(\omega)_+ = \int_0^{\infty}\dd\tau\ \ee^{\ii\omega\tau} G_{c,x}(\tau).
\end{equation}
The Langevin (fluctuation) operators are expressed as
\begin{subequations}
\begin{align}
\oFin_c(t)
& = \sum_m \kappa_m \oalpha_m(t_0) \ee^{-\ii\varOmega^c_m(t-t_0)}
= \sum_m \kappa_m \oalphain_m \ee^{-\ii\varOmega^c_m t}, \\
\oFin_x(\omega)
& = \sum_m \gamma_m \obeta_m(t_0) \ee^{-\ii\varOmega^x_m(t-t_0)}
= \sum_m \gamma_m \obetain_m \ee^{-\ii\varOmega^x_m t},
\end{align}
\end{subequations}
where $t_0 \rightarrow -\infty$ is the switch-on time of system-reservoir interaction,
and $\oalphain_m$ and $\obetain_m$ are the input operators.
Their Fourier transforms are derived as
\begin{subequations}
\begin{align}
\oFin_c(\omega)
& = \int_{-\infty}^{\infty}\dd t\ \ee^{\ii\omega t} \oFin_c(t)
= 2\pi\theta(\omega)\kappa(\omega)\oalphain(\omega), \\
\oFin_x(\omega)
& = \int_{-\infty}^{\infty}\dd t\ \ee^{\ii\omega t} \oFin_x(t)
= 2\pi\theta(\omega)\gamma(\omega)\obetain(\omega).
\end{align}
\end{subequations}
Here, the reservoir states are rewritten in continuous form as
in Eqs.~\eqref{eq:discrete2continuous}.
These fields $\oFin_{c,x}(\omega)$ are interpreted as the input fields,
and they cannot be defined for negative frequency $\omega < 0$,
because the reserver states are distributed only for positive frequencies
$\varOmega_j^{c,x} > 0$.

According to the input-output formalism \cite{Ciuti2006PRA},
the output photonic field (photonic reservoir field at time $t_1 \rightarrow \infty$)
is represented as
\begin{equation}
\oalphaout(\omega>0) = \oalphain(\omega)
+ \kappa(\omega)^*\oa(\omega).
\end{equation}
As discussed by Ciuti and Carusotto \cite{Ciuti2006PRA},
we get the vacuum output for vacuum input.
However, the $\oHz$ system is actually excited by the vacuum reservoirs as
$\braket{\opd_j(\omega)\op_k} \neq 0$
and $\braket{\op_j(\omega)\op_k} \neq 0$,
which can be easily verified from the Langevin equations
\eqref{eq:Langevin_ex_ph}.
In the master-equation formalism discussed in the previous section,
the $\oHz$ system is also excited,
but the vacuum output is not obtained for the vacuum input.
Then, there is a discrepancy between the two approaches
at least under the Born approximation.
Instead, in the input-output formalism,
we also suppose the squeezed and correlated reservoirs
discussed in Secs.~\ref{app:diag_whole} and \ref{sec:master}.

According to the standard input-output formalism,
the output photonic field is represented as
\begin{subequations} \label{eq:input-output-F} 
\begin{align}
\oFout_c(t) & = \oFin_c(t) + \int_{-\infty}^{\infty}\dd t'\ \Gc(t-t') \oa(t'), \\
\oFout_c(\omega) & = \oFin_c(\omega) + \Gc(\omega) \oa(\omega).
\end{align}
\end{subequations}
This expression does not violate the causality ($\oFout_c(t)$ can be affected
by $\oa(t'>t)$) as discussed in Sec.~\ref{sec:master}.
From this input-output relation and the Langevin equations,
the output photonic field is eventually represented
by the input fields $\oFin_{c,x}(\omega)$.
For discussing the output from the cavity,
we have to suppose the correlation of input operators  $\{\oFin_{c,x}(\omega)\}$.
Here, we consider that the $\oHz$ system is in the ground state,
and the correlation of input operators are also
supposed as shown in Eq.~\eqref{eq:<Fz(t)Fz>}:
\begin{equation} \label{eq:<Fzcx(w)Fzcx>} 
\braketR{\ovFin_{cx}(\omega)\ovFin_{cx}{}^{\text{T}}}
= \braketR{\ovFin_{cx}(\omega)_+\ovFin_{cx}{}^{\text{T}}}
+ \braketR{\ovFin_{cx}(\omega)_-\ovFin_{cx}{}^{\text{T}}},
\end{equation}
\begin{widetext}
\begin{equation}
\braketR{\ovFin_{cx}(\omega)_+\ovFin_{cx}{}^{\text{T}}}
= \begin{pmatrix}
\Gc(\omega)_+\braketg{\oa\oad} & \Gc(\omega)_+\braketg{\oa\obd} & 
\Gc(\omega)_+\braketg{\oa\oa} & \Gc(\omega)_+\braketg{\oa\ob} \\
\Gx(\omega)_+\braketg{\ob\oad} & \Gx(\omega)_+\braketg{\ob\obd} & 
\Gx(\omega)_+\braketg{\ob\oa} & \Gx(\omega)_+\braketg{\ob\ob} \\
\Gc(-\omega)_+^*\braketg{\oad\oad} & \Gc(-\omega)_+^*\braketg{\oad\obd} & 
\Gc(-\omega)_+^*\braketg{\oad\oa} & \Gc(-\omega)_+^*\braketg{\oad\ob} \\
\Gx(-\omega)_+^*\braketg{\obd\oad} & \Gx(-\omega)_+^*\braketg{\obd\obd} & 
\Gx(-\omega)_+^*\braketg{\obd\oa} & \Gx(-\omega)_+^*\braketg{\obd\ob}
\end{pmatrix},
\end{equation}
\begin{equation}
\braketR{\ovFin_{cx}(\omega)_-\ovFin_{cx}{}^{\text{T}}}
= \begin{pmatrix}
\Gc(\omega)_+^*\braketg{\oa\oad} & \Gx(\omega)_+^*\braketg{\oa\obd} & 
\Gc(-\omega)_+\braketg{\oa\oa} & \Gx(-\omega)_+\braketg{\oa\ob} \\
\Gc(\omega)_+^*\braketg{\ob\oad} & \Gx(\omega)_+^*\braketg{\ob\obd} & 
\Gc(-\omega)_+\braketg{\ob\oa} & \Gx(-\omega)_+\braketg{\ob\ob} \\
\Gc(\omega)_+^*\braketg{\oad\oad} & \Gx(\omega)_+^*\braketg{\oad\obd} & 
\Gc(-\omega)_+\braketg{\oad\oa} & \Gx(-\omega)_+\braketg{\oad\ob} \\
\Gc(\omega)_+^*\braketg{\obd\oad} & \Gx(\omega)_+^*\braketg{\obd\obd} & 
\Gc(-\omega)_+\braketg{\obd\oa} & \Gx(-\omega)_+\braketg{\obd\ob}
\end{pmatrix},
\end{equation}
\end{widetext}
where $\ovFin_{cx}(\omega) = [\oFin_c(\omega), \oFin_x(\omega), \oFin_c(-\omega)^{\dagger}, \oFin_x(-\omega)^{\dagger}]^{\text{T}}$.
Precisely speaking the expectation values such as $\braketg{\oad\oa}$
should be slightly modified depending on $\omega$
as discussed in App.~\ref{app:ordering_input-output}.
Assuming this input correlation, the system is certainly in the ground state
$\braket{\opd_j(\omega)\op_k} = \braket{\op_j(\omega)\op_k} = 0$ ($j,k=L,U)$,
and we also obtain no photon detection
$\braket{:\oFout_c(\omega)\oFoutd_c:} = 0$
and $\braket{:\oFout_c(\omega)\oFout_c:} = 0$
by normal- and time-ordering the operators in the polariton base.
The detailed calculation is shown in App.~\ref{app:ordering_input-output}.

In this way, when we suppose the squeezed and correlated reservoirs 
represented in Eqs.~\eqref{eq:<Fz(t)Fz>} and \eqref{eq:<Fzcx(w)Fzcx>},
the $\oHz$ system certainly decays to its ground state $\ketg$
and no photon is detected outside the cavity
in both formalisms of master equation and input-output relation.
In contrast, when we suppose the vacuum reservoirs,
different results are obtained in the two formalisms.

\section{Discussion} \label{sec:discussion}
As discussed in the previous sections,
when we consider the squeezed and correlated reservoirs
instead of the vacuum ones,
both master-equation and input-output formalisms certainly guarantee
the decay of the $\oHz$ system to its ground state
and show no photon detection, even though we do not use the RWA
on system-reservoir coupling. 
The supposed reservoir state is approximately equivalent to 
the one realized in the ground state of the whole system:
$\orhoR \simeq \Tr_R\{\ketgg\bragg\}$.
On the other hand, if we suppose the vacuum reservoirs in excitation-photon base,
in the absence of the RWA on system-reservoir coupling,
the $\oHz$ system is excited by the coupling with the reservoirs
as seen in Fig.~\ref{fig:1}.
We have to determine the reservoir state supposed in the master equation and input-output formalism,
according to the situation how the system and reservoirs start to couple.
If we initially prepare the vacuum reservoirs and switch on the system-reservoir
coupling,
the reservoirs approximately remain in the vacuum state even after the switch-on,
and the system does not decay to its ground state but to a steady state
excited by the vacuum reservoirs as seen in Fig.~\ref{fig:1}.
This is because the $\oHz$ system is excited
when virtual photons escape to the vacuum reservoirs.
In order to avoid it, the $\oHz$ system and the reservoirs should be balanced
as realized in the ground state $\ketgg$ of the whole system,
and we should suppose $\orhoR \simeq \Tr_R\{\ketgg\bragg\}$ in such situation.
If we consider that the system and reservoirs are already coupled
and the whole system is in the ground state,
when we excite the $\oHz$ system to an excited state,
the system certainly decays to its ground state as seen in Fig.~\ref{fig:2}.

If the reservoirs are quite large and the whole system cannot be in a steady state,
we should suppose the former situation.
When the temperatures of the reservoirs are low enough
and the vacuum input from the reservoirs to the system is supposed,
the $\oHz$ system in principle does not decay to its ground state.
Although the vacuum output is obtained according to the input-output formalism
\cite{Ciuti2006PRA}, it is not by the master equation.
The energy is conserved in the input-output formalism,
but it seems not in the master equation.
This is because the dynamics of focusing system
and the output are discussed in the sense of perturbation theory
in the formalism of master equation.
In this way, when we suppose the vacuum input,
we should pay attention to the difference of the two formalisms
(at least under the Born approximation).
In order to avoid this discrepancy,
we should use the RWA on system-reservoir coupling,
although the quantum fluctuation of reservoirs is diminished
in such treatment.

On the other hand, if we can define relatively small reservoirs
which weakly couple with a large external system with low enough temperature,
the small reservoirs and the $\oHz$ system can decay to
the ground state $\ketgg$ of the coupled system.
In such situation, we can suppose $\orhoR \simeq \Tr_R\{\ketgg\bragg\}$,
and it guarantees the decay of $\oHz$ system to its ground state $\ketg$
and gives no photon detection in the small reservoir
as discussed in the previous sections.
This result is obtained in both formalisms of master equation
and of input-output relation
in contrast to supposing the vacuum reservoirs in excitation-photon base.


As discussed in Ref.~\cite{Beaudoin2011PRA},
by performing the RWA on system-reservoir coupling,
we can simply suppose the vacuum reservoirs in excitation-photon base,
and the master equation is reduced to the standard Lindblad form.
The simplified master equation is derived as follows.
Whereas the system-reservoir coupling is originally represented as
Eq.~\eqref{eq:oHint-polariton}, here we perform the pre-trace RWA
\cite{Fleming2010JPA} as
\begin{equation} \label{eq:oHint-pre-RWA} 
\oHint \simeq \ii\hbar\sum_{j=L,U}(w_j^*\oFd_c\op_j - w_j\opd_j\oF_c
 + x_j^*\oFd_x\op_j - x_j\opd_j\oF_x),
\end{equation}
where the counter-rotating terms are neglected in the polariton base
not in the excitation-photon base.
Then, when we suppose the vacuum reservoirs in the excitation-photon base
as in Eqs.~\eqref{eq:<FzFz>-vacuum},
the master equation is derived under the Born approximation as
\begin{align}
\ddt{}\orho
& = \frac{1}{\ii\hbar}[ \oHz, \orho ]
\nonumber \\ & \quad
  + \sum_{j,k=L,U} \int_{t_0}^t\dd t'\ \{G_c(t-t')w_jw_k^*+G_x(t-t')x_jx_k^*\}
\nonumber \\ & \quad \times
    \left[ \oUz(t-t')[\op_k\orho(t')], \opd_j\right]
+ \Hc
\end{align}
Further, by neglecting the fast oscillating terms
$\op_k\opd_j\ee^{-\ii(\omega_k-\omega_j)t}$ for $j \neq k$
(called the post-trace RWA \cite{Fleming2010JPA}),
we finally get the simplified master equation under the Markov approximation as
\begin{align}
\ddt{}\orho
& = \frac{1}{\ii\hbar}[ \oHz, \orho ]
\nonumber \\ & \quad
  + \sum_{j=L,U} \frac{\varGamma_c|w_j|^2+\varGamma_x|x_j|^2}{2}
    (2\op_j\orho\opd_j - \opd_j\op_j\orho - \orho\opd_j\op_j),
\end{align}
where the memory kernels are approximated
as $G_{\mu}(t) = \varGamma_\mu\delta(t)$ for simplicity
(there remains the Lamb-shift terms in general \cite{Beaudoin2011PRA}).
If the system-reservoir coupling is expressed in the Hermitian form as
$\ii\hbar(\oa\pm\oad)(\oFd_c\mp\oF_c)$ [$\ii\hbar(\ob\pm\obd)(\oFd_x\mp\oF_x)$],
the above master equation is simplify rewritten
by replacing $w_j$ [$x_j$] by $w_j \mp y_j$ [$x_j \mp z_j$].
Even in such case, the simplified master equation
is represented in the Lindblad form.
From Eq.~\eqref{eq:oHint-pre-RWA}, the input-output relation is obtained as
\begin{align}
\oFout_c & = \oFin_c + \varGamma_c \sum_j w_j^*\op_j.
\end{align}
Since the above master equation is reduced to the standard form
owing to the pre-trace and post-trace RWAs,
we consider that the correlation of input operator $\oFin_c$
is equivalent to that of $\oFz_c$ supposed in the master equation:
$\braketR{\oFin_c(t)\oFind_c} = G_c(t)$
and $\braketR{\oFind_c(t)\oFin_c} = \braketR{\oFin_c(t)\oFin_c} = 0$.
Then, the correlation of the output can be calculated
as discussed in Ref.~\cite{Ridolfo2012arXiv}.

However, in this approach,
the photonic and excitonic reservoirs are supposed
to be in the vacuum state under the RWA on system-reservoir coupling,
although the polariton system does not decay to its ground state $\ketg$ in general
without the RWA.
In other words,
the quantum statistics of reservoirs fields are diminished by the RWA,
although the reservoirs are originally squeezed and correlated.
In contrast, in the present paper, the master equation and input-output formalism
are discussed based on the squeezed and correlated reservoirs.
The master equation certainly guarantees the decay of $\oHz$ system
to its ground state,
and in both formalisms
no photon is detected when the $\oHz$ system is in the ground state.
Under the Markov approximation
the master equation \eqref{eq:master_Born_diss} is reduced to
\begin{align}
\ddt{}\orho
& = \frac{1}{\ii\hbar}[\oHz,\orho]
+ \sum_{j,k=L,U} \frac{\varGamma_{j,k}}{2}
  (2\op_j\orho\opd_k - \opd_k\op_j\orho - \orho\opd_k\op_j)
\nonumber \\ & \quad
+ \sum_{j,k=L,U} \left\{\frac{K_{j,k}}{2}
  (\op_j\orho\op_k - \op_k\op_j\orho) + \Hc \right\}.
\label{eq:master_Born-Markov_diss_simple} 
\end{align}
The coefficients $\varGamma_{j,k}$ and $K_{j,k}$
can be calculated from the supposed free field-correlation
in Eqs.~\eqref{eq:<Fz(t)Fz>}.
This does not have the Lindblad form,
but certainly guarantees the decay to the ground state $\ketg$
as seen in Fig.~\ref{fig:2}.

In the standard theory \cite{gardiner04,Breuer2006,walls08}
and also in the discussion of Refs.~\cite{Beaudoin2011PRA,Ridolfo2012arXiv},
the master equation and the input-output relation are sometimes used together
and the correlation of input $\oFin_c$ is supposed to be equal
to that of free field $\oFz_c$ given in the master equation.
However, 
the formalism of master equation is discussed in the sense of perturbation theory.
Since the reservoirs are large enough compared to the $\oHz$ system,
the input correlation is not strongly modified
and constantly given in the master equation.
On the other hand, the output is a perturbation of the reservoirs
as a result of the system-reservoir coupling.
The correlation of $\oFz_c$ can be in general different on input and output sides.
Actually, when we suppose the vacuum reservoirs in excitation-photon base,
the self-correlation of $\oFz_c$ on output side is not in vacuum,
which is calculated by Eq.~\eqref{eq:<Fz(t)Fz>}.
In order to get the same correlation for input and output sides,
we have to consider the squeezed and correlated input
$\orhoR \simeq \Tr_R\{\ketgg\bragg\}$.
If we want to reduce this complicated formalism into the standard one,
we have to perform the RWA on system-reservoir coupling
\cite{Beaudoin2011PRA,Ridolfo2012arXiv}.

If we already know that the free field $\oFz_c$ does not contribute
to the observables, we can simply use the RWA on system-reservoir coupling
\cite{Beaudoin2011PRA,Ridolfo2012arXiv}.
For example, the second-order correlation functions
under a resonant excitation can be calculated
as discussed in Ref.~\cite{Ridolfo2012arXiv}.
However, when we discuss squeezing of the emission,
the interference between free field $\oFz_c$ and cavity contribution
is important, and the quantum fluctuation of $\oFz_c$ should not
be destroyed by the RWA on system-reservoir coupling.
If the cavity system has an optical nonlinearity or embeds ensemble of atoms,
we have to treat the Langevin equations perturbatively
or the master equation might be appropriate to treat such systems.
When we discuss the emission (or lasing) from such complex systems
under incoherent excitation, it is difficult to evaluate the validity
of the RWA on system-reservoir coupling,
and we should suppose the squeezed and correlated reservoirs
realized in the ground state of the whole system.
This kind of approach should give us natural results
in the calculation of dissipation and detection of output.

\section{Summary} \label{sec:summary}
We have derived the master equation, Langevin equations, and input-output
relation for dissipative polariton system in the ultrastrong light-matter coupling regime.
The correlation of reservoir free fields are required for calculating
not only the dynamics of the system but also the photon emission from the polariton system.
When we suppose the vacuum reservoirs, the polariton system is excited in general.
Although the vacuum output is obtained for the vacuum input in the input-output formalism,
it is not obtained in the master-equation approach under the Born approximation.
In order to avoid this discrepancy, we have to perform the RWA
on system-reservoir coupling, although it diminishes the 
quantum statistics of the reservoirs.
In order to describe the dissipation in the ultrastrong coupling regime
without the RWA on system-reservoir coupling,
we have considered the correlation functions of the photonic and excitonic free fields
that are squeezed and correlated with each other
and realized in the ground state $\ketgg$ of the whole system:
$\orhoR \simeq \Tr_S\{\ketgg\bragg\}$.
In the formalism of master equation,
the supposed correlation certainly guarantees the decay
of the polariton system to its original ground state $\ketg$.
In the ground state,
we have also verified no photon detection as the output from the cavity.
Even in the formalism of Langevin equations and input-output relation,
we also get no photon detection
by considering the squeezed and correlated reservoirs.
This reservoir state is also consistent
to the analysis of the ground state of the whole system
by the Fano-type diagonalization technique.
At least when the polariton system is dissipative and is in the ground state,
the three approaches, master equation, input-output formalism,
and Fano-type diagonalization give the same result,
in contrast to supposing the vacuum reservoirs.
The case in the presence of excitation to the system will be discussed
in the future.

\begin{acknowledgments}
The authors thank to Cristiano Ciuti, Howard Carmichael, Salvatore Savasta, Pierre Nataf, Kenji Kamide, Makoto Yamaguchi, and Tatsuro Yuge
for fruitful discussion.
This work was supported by KAKENHI (No.~20104008 and No.~24-632)
and the JSPS through its FIRST Program.
\end{acknowledgments}

\appendix
\section{Diagonalization of photonic and excitonic parts} \label{app:diag_photon}
In order to diagonalize the whole Hamiltonian $\oH = \oHz + \oHint + \oHR$,
first of all, we diagonalize the photonic part, Eq.~\eqref{eq:oHa}.
Here, we rewrite the reservoir fields
from discrete to continuous form as
\begin{subequations} \label{eq:discrete2continuous} 
\begin{align}
\oalpha_m & \rightarrow \oalpha(\varOmega^c_m) / \sqrt{\DOSc(\varOmega^c_m)}, \\
\obeta_m & \rightarrow \obeta(\varOmega^x_m) / \sqrt{\DOSx(\varOmega^x_m)}, \\
\kappa_m & \rightarrow \kappa(\varOmega^c_m) / \sqrt{\DOSc(\varOmega^c_m)}, \\
\gamma_m & \rightarrow \gamma(\varOmega^x_m) / \sqrt{\DOSx(\varOmega^x_m)}.
\end{align}
\end{subequations}
where $\DOSc(\omega)$ and $\DOSx(\omega)$ are densities of states of photonic
and excitonic reservoirs, respectively.
The new reservoir operators satisfy
$[ \oalpha(\omega), \oalphad(\omega') ]
= [ \obeta(\omega), \obetad(\omega') ] = \delta(\omega-\omega')$.
The photonic Hamiltonian is rewritten as
\begin{align}
\oHa & = \hbar\wc\oad\oa
+ \int_0^{\infty}\dd\omega\ \hbar\omega\oalphad(\omega)\oalpha(\omega)
\nonumber \\ & \quad
+ \ii\hbar\int_0^{\infty}\dd\omega \left[ \kappa(\omega)^* \oalphad(\omega) \oa - \oad \oalpha(\omega) \kappa(\omega) \right].
\end{align}
This kind of Hamiltonian can be diagonalized by the Fano-type technique
\cite{Fano1961PR,Barnett1988OC,huttner92}
by introducing an operator for eigen frequency $\omega$ as
\begin{equation} \label{eq:ooa} 
\ooa(\omega) = \uc(\omega) \oa + \int_0^{\infty}\dd\omega'\ \vc(\omega,\omega') \oalpha(\omega').
\end{equation}
Once this operator satisfies Eq.~\eqref{eq:[c,H]=hwc},
we can diagonalize the photonic Hamiltonian as in Eq.~\eqref{eq:oHa}.
Further, $\ooa(\omega)$ should be normalized as
\begin{align}
& \left[ \ooa(\omega), \ooad(\omega') \right] \nonumber \\
& = \uc(\omega)\uc(\omega')^*
+ \int_0^{\infty}\dd\omega''\ \vc(\omega,\omega'')\vc(\omega',\omega'')^*
\nonumber \\
& = \delta(\omega-\omega').
\end{align}

The coefficient functions $\uc(\omega)$ and $\vc(\omega,\omega')$
are determined as follows.
From Eq.~\eqref{eq:[c,H]=hwc}, we get
\begin{subequations}
\begin{align}
\omega\uc(\omega)
& = \wc\uc(\omega) + \ii\int_0^{\infty}\dd\omega'\ \vc(\omega,\omega')\kappa(\omega'),
\label{eq:omega_alpha=} \\ 
\omega\vc(\omega,\omega')
& = \omega'\vc(\omega,\omega') - \ii\kappa(\omega')\uc(\omega).
\end{align}
\end{subequations}
From the second equation, $\vc(\omega,\omega')$ is expressed as
\begin{subequations} \label{eq:vc} 
\begin{align}
& \vc(\omega,\omega') \nonumber \\
& = - \ii\kappa(\omega')\frac{\uc(\omega)}{\omega-\omega'}, \\
& = - \ii\kappa(\omega')\left\{ \frac{\PP}{\omega-\omega'} + \psi(\omega)\delta(\omega-\omega') \right\}\uc(\omega), \\
& = - \ii\kappa(\omega')\left\{ \frac{1}{\omega-\omega'-\ii0^+} + [\psi(\omega)-\ii\pi]\delta(\omega-\omega') \right\}\uc(\omega),
\end{align}
\end{subequations}
where $\PP$ means the principal value integral
and function $\psi(\omega)$ is introduced for the following calculation.
The expression of $\psi(\omega)$ is determined
by substituting the second or third equation into
Eq.~\eqref{eq:omega_alpha=} as
\begin{subequations}
\begin{align}
\psi(\omega) & = \frac{1}{|\kappa(\omega)|^2}\left\{ \omega - \wc - \PP\int_0^{\infty}\dd\omega'\ \frac{|\kappa(\omega')|^2}{\omega-\omega'} \right\}, \\
\psi(\omega) - \ii\pi & = \frac{1}{|\kappa(\omega)|^2}\left\{ \omega - \wc - \int_0^{\infty}\dd\omega'\ \frac{|\kappa(\omega')|^2}{\omega-\omega'-\ii0^+} \right\}.
\end{align}
\end{subequations}
On the other hand, the expression of $\uc(\omega)$ is determined by the normalization
condition, Eq.~\eqref{eq:[oc,ocd]}.
The commutator is derived as
\begin{equation}
\left[ \ooa(\omega), \ooad(\omega') \right]
= \uc(\omega)\uc^*(\omega') [ \psi(\omega) - \ii\pi ]
  [ \psi(\omega) + \ii\pi ] \delta(\omega-\omega').
\end{equation}
Then, we get
\begin{equation} \label{eq:uc} 
\uc(\omega) = \frac{1}{\psi(\omega)-\ii\pi}
= \frac{|\kappa(\omega)|^2}{\omega-\wc \zeta(\omega)},
\end{equation}
where
\begin{equation}
\zeta(\omega) = 1 - \frac{1}{\wc} \int_0^{\infty}\dd\omega'\ \frac{|\kappa(\omega')|^2}{\omega'-\omega+\ii0^+}.
\end{equation}
Using the diagonalized operator $\ooa(\omega)$, the original ones are represented as
\begin{subequations} \label{eq:ooa2oa} 
\begin{align}
\oa & = \int_0^{\infty}\dd\omega\ \uc(\omega)^* \ooa(\omega), \\
\oalpha(\omega) & = \int_0^{\infty}\dd\omega'\ \vc(\omega',\omega)^* \ooa(\omega').
\end{align}
\end{subequations}

In the same manner, 
we can also diagonalize the excitonic Hamiltonian
as in Eq.~\eqref{eq:oHb}.
The eigen operator is represented as
\begin{equation} \label{eq:oob} 
\oob(\omega) = \ux(\omega) \ob + \int_0^{\infty}\dd\omega'\ \vx(\omega,\omega') \obeta(\omega').
\end{equation}
The coefficient functions are determined in the same manner
by replacing $\wc$ and $\kappa(\omega)$ with $\wx$ and $\gamma(\omega)$,
respectively.
The excitations and excitonic reservoir field are represented as
\begin{subequations} \label{eq:oob2ob} 
\begin{align}
\ob & = \int_0^{\infty}\dd\omega\ \ux(\omega)^* \oob(\omega), \\
\obeta(\omega) & = \int_0^{\infty}\dd\omega'\ \vx(\omega',\omega)^* \oob(\omega').
\end{align}
\end{subequations}

\section{Correlation of free field}\label{app:corr_system_reservoir}
By using the technique in Ref.~\cite{Carmichael1987JOSAB},
here we calculate the correlation between the free field $\oFz_{c,x}(t)$
and system operators $\oa$ and $\ob$.
Further, we also derive the self-correlation of $\oFz_{c,x}(t)$.

First, let's calculate $\braket{\oS(t)\oFz_c(t+\tau)}$ for arbitrary system operator $\oS$ and $\tau > 0$.
From the Langevin equation \eqref{eq:motion-a-b},
the free field is represented as
\begin{equation} \label{eq:iFz=} 
\oFz_c(t) = - \ddt{}\oa(t) + \frac{1}{\ii\hbar}\left[\oa(t), \oHz\right] - \int_{t_0}^t\dd t'\ \Gc(t-t') \oa(t'),
\end{equation}
then we get
\begin{align} \label{eq:<SF0>=} 
& \braket{\oS(t)\oFz_c(t+\tau)} \nonumber \\
& = - \ddtau{}\braket{\oS(t)\oa(t+\tau)}
  + \frac{1}{\ii\hbar}\braket{\oS(t)[\oa, \oHz](t+\tau)}
\nonumber \\ & \quad
- \int_{t_0}^{t+\tau}\dd t'\ \Gc(t+\tau-t') \braket{\oS(t)\oa(t')}.
\end{align}
The correlation functions of system operators
appearing on the right hand side
can be calculated by the master equation \eqref{eq:master_Born}.
By using the quantum regression theorem \eqref{eq:QRT},
the first term of Eq.~\eqref{eq:<SF0>=} is rewritten as
\begin{align}
& \ddtau{}\braket{\oS(t)\oa(t+\tau)} \nonumber \\
& = \Tr\{\oa\oL[\oU(\tau)[\orho(t)\oS]]\} \nonumber \\
& = \frac{1}{\ii\hbar} \braket{\oS(t)[\oa, \oHz](t+\tau)}
  + \Tr\{\oa\oLdiss[\oU(\tau)[\orho(t)\oS]]\},
\end{align}
and its last term is also written as
\begin{align}
\Tr\{\oa\oLdiss[\oU(\tau)[\orho(t)\oS]]\}
= - \int_{t}^{t+\tau}\dd t'\
\braket{\oS(t)\oa(t')} G(t-t').
\end{align}
Substituting these two equations into Eq.~\eqref{eq:<SF0>=},
we get Eq.~\eqref{eq:<S(t)Fz(t+tau)>}.

Next, let's consider $\braket{\oFz_c(t+\tau)\oS(t)}$ for $\tau > 0$.
From Eq.~\eqref{eq:iFz=}, we get
\begin{align} \label{eq:<F0S>=} 
& \braket{\oFz_c(t+\tau)\oS(t)} \nonumber \\
& = - \ddtau{}\braket{\oa(t+\tau)\oS(t)}
  + \frac{1}{\ii\hbar}\braket{[\oa(t+\tau), \oHz]\oS(t)}
\nonumber \\ & \quad
  - \int_{t_0}^{t+\tau}\dd t'\ \Gc(t+\tau-t') \braket{\oa(t')\oS(t)}.
\end{align}
In the same manner as the above calculation, the first term is rewritten as
\begin{align}
& \ddtau{}\braket{\oa(t+\tau)\oS(t)} \nonumber \\
& = \frac{1}{\ii\hbar} \braket{[\oa, \oHz](t+\tau)\oS(t)}
  - \int_{t_0}^{t+\tau}\dd t'\ \Gc(t+\tau-t') \braket{\oa(t')\oS(t)},
\end{align}
and we get Eq.~\eqref{eq:<Fz(t+tau)S(t)>}.

The next is $\braket{\oS(t+\tau)\oFz_c(t)}$ for $\tau > 0$.
From Eq.~\eqref{eq:iFz=}, we get
\begin{align}
& \braket{\oS(t+\tau)\oFz_c(t)} \nonumber \\
& = - \braket{\oS(t+\tau)\ddt{}\oa(t)}
  + \frac{1}{\ii\hbar}\braket{\oS(t+\tau)[\oa, \oHz](t)}
\nonumber \\ & \quad
- \int_{t_0}^{t}\dd t'\ \Gc(t-t') \braket{\oS(t+\tau)\oa(t')}.
\label{eq:<S(t+tau)F0(t)>=} 
\end{align}
As shown in Ref.~\cite{Carmichael1987JOSAB}, the first term is represented as
\begin{align}
& \braket{\oS(t+\tau)\ddt{}\oa(t)} \nonumber \\
& = \frac{1}{\ii\hbar}\braket{\oS(t+\tau)[\oa, \oHz](t)}
\nonumber \\ & \quad
  + \Tr\{\oS\oU(\tau)[\oa\oLdiss[\orho(t)]-\oLdiss[\oa\orho(t)]]\},
\end{align}
and then we get Eqs.~\eqref{eq:<S(t)Fz>}.
The second equation is also derived in the same manner,
and similar expressions are obtained also for $\oFz_x$

Since the free field $\oFz_c(t)$ is expressed as in Eq.~\eqref{eq:iFz=},
the self-correlation is represented as
\begin{widetext}
\begin{multline} \label{eq:<FzFzd>} 
\braket{\oFz_c(t)\oFzd_c(\tau)}
= \ddt{}\ddtau{}\braket{\oa(t)\oad(\tau)}
- \frac{1}{\ii\hbar}\ddt{}\braket{\oa(t)[\oad, \oHz](\tau)}
+ \int_{t_0}^{\tau}\dd t'\ \Gc^*(\tau-t') \ddt{}\braket{\oa(t)\oad(t')} \\
- \frac{1}{\ii\hbar}\ddtau{}\braket{[\oa, \oHz](t)\oad(\tau)}
+ \frac{1}{(\ii\hbar)^2}\braket{[\oa, \oHz](t)[\oad, \oHz](\tau)}
- \frac{1}{\ii\hbar}\int_{t_0}^{\tau}\dd t'\ \Gc^*(\tau-t') \braket{[\oa, \oHz](t)\oad(t')} \\
+ \int_{t_0}^{t}\dd t''\ \Gc(t-t'') \ddtau{}\braket{\oa(t'')\oad(\tau)}
- \frac{1}{\ii\hbar}\int_{t_0}^{t}\dd t''\ \Gc(t-t'')\braket{\oa(t'')[\oad, \oHz](\tau)} \\
+ \int_{t_0}^{t}\dd t'' \int_{t_0}^{\tau}\dd t'\ \Gc(t-t'') \Gc^*(\tau-t') \braket{\oa(t'')\oad(t')}.
\end{multline}
In the same manner as discussed above,
the first term is rewritten for $\tau > t$ as
\begin{align}
\ddt{}\ddtau{}\braket{\oa(t)\oad(\tau)}
& = \ddt{}\ddtau{} \Tr\{\oad\oU(\tau-t)[\orho(t)\oa]\} \nonumber \\
& = \frac{1}{\ii\hbar} \ddt{} \braket{\oa(t)[\oad,\oHz](\tau)}
  - \ddt{}\int_t^{\tau}\dd t'\ \braket{\oa(t)\oad(t')} \Gc^*(\tau-t') \nonumber \\
& = \frac{1}{\ii\hbar} \ddt{} \braket{\oa(t)[\oad,\oHz](\tau)}
  - \int_t^{\tau}\dd t'\ \ddt{}\braket{\oa(t)\oad(t')} \Gc^*(\tau-t')
  + \Gc^*(\tau-t) \braket{\oa(t)\oad(t)}.
\end{align}
Rewriting 4th and 7th terms in Eq.~\eqref{eq:<FzFzd>},
the self-correlation is reduced to
\begin{multline}
\braket{\oFz_c(t)\oFzd_c(\tau)}
= \Gc^*(\tau-t) \braket{\oa(t)\oad(t)}
+ \int_{t_0}^{t}\dd t'\ \Gc^*(\tau-t') \ddt{}\braket{\oa(t)\oad(t')} \\
- \frac{1}{\ii\hbar} \int_{t_0}^{t}\dd t'\ \Gc^*(\tau-t') \braket{[\oa,\oHz](t)\oad(t')}
+ \int_{t_0}^{t}\dd t'' \int_{t_0}^{t''}\dd t'\ \Gc(t-t'') \Gc^*(\tau-t') \braket{\oa(t'')\oad(t')}.
\end{multline}
Rewriting the second term, we finally get
\begin{align}
\braket{\oFz_c(t)\oFzd_c(\tau)}
& = \Gc^*(\tau-t) \braket{\oa(t)\oad(t)}
- \int_{t_0}^{t}\dd t' \int_{t'}^{t}\dd t''\ \Gc(t-t'') \Gc^*(\tau-t') \braket{\oa(t'')\oad(t')}
\nonumber \\ & \quad
+ \int_{t_0}^{t}\dd t'' \int_{t_0}^{t''}\dd t'\ \Gc(t-t'') \Gc^*(\tau-t') \braket{\oa(t'')\oad(t')} \nonumber \\
& = \Gc^*(\tau-t) \braket{\oa(t)\oad(t)}.
\end{align}
In the same manner, we finally get Eqs.~\eqref{eq:<Fz(t)Fz>}.
\end{widetext}

\section{Calculation of observables in master-equation approach} \label{app:ordering_master}
When we detect photons emitted from the cavity,
the observables by photon detectors should be calculated
by normal- and time-ordering the photon operators
\cite{Carmichael1987JOSAB,gardiner04,walls08}.
In the present case, the ordering should be performed
in the polariton basis, which really represents the eigen states of the system.
Here, we divide the photon operator $\oa$ into the lowering parts $\oa_{\low}$
and raising part $\oa_{\rai}$ as
\begin{subequations}
\begin{align}
\oa_{\low} & = w_L^* \op_L + w_U^* \op_U, \\
\oa_{\rai} & = - y_L \opd_L - y_U \opd_U.
\end{align}
\end{subequations}
Since the system-reservoir coupling is expressed
as in Eq.~\eqref{eq:oHint-polariton},
the photonic free field $\oFz_c$ is also divided as
\begin{subequations}
\begin{align}
\oFz_{c\low} & = w_L^* \oFz_L + w_U^* \oFz_U, \\
\oFz_{c\rai} & = - y_L \oFzd_L - y_U \oFzd_U,
\end{align}
\end{subequations}
and then the output field \eqref{eq:def-iFout} is rewritten as
\begin{equation}
\oFout_c = \oFout_{c\low} + \oFout_{c\rai},
\end{equation}
\begin{subequations}
\begin{align}
\oFout_{c\low}(t) & = \oFz_{c\low}(t) + \int_{t_0}^{t+z_0/c}\dd t'\ \Gc(t-t') \oa_{\low}(t'), \\
\oFout_{c\rai}(t) & = \oFz_{c\rai}(t) + \int_{t_0}^{t+z_0/c}\dd t'\ \Gc(t-t') \oa_{\rai}(t').
\end{align}
\end{subequations}
The equation of motion of $\oa_{\low\rai}$ is derived as
\begin{align}
\ddt{}\oa_{\low\rai}(t)
& = \frac{1}{\ii\hbar}[\oa_{\low\rai},\oHz](t)
\nonumber \\ & \quad
  - \int_{t_0}^t\dd t'\ \Gc(t-t') \oa_{\low\rai}(t') - \oFz_{c\low\rai}(t).
\end{align}
Then, the correlation between $\oa_{\low\rai}$ and $\oFz_{c\low\rai}$
is also derived in the same form as
Eqs.~\eqref{eq:<SFz(t)>} and \eqref{eq:<S(t)Fz>},
and the self correlation of $\oFz_{c\low\rai}$
also has the same form as Eq.~\eqref{eq:<Fz(t)Fz>}.

The output field measured by photon detectors outside the cavity
is calculated by normal- and time-ordering
the operators.
The emission spectrum (number of photons) in a steady state is expressed as
\begin{align}
& \braket{:\oFout_c(\omega)\oFoutd_c:}
\nonumber \\ &
= \braket{:\oFz_c(\omega)\oFzd_c:}
+ \Gc(\omega)\braket{:\oa(\omega)\oFzd_c:}
\nonumber \\ & \quad
+ \Gc(\omega)^*\braket{:\oFz_c(\omega)\oad:}
+ |\Gc(\omega)|^2\braket{:\oa(\omega)\oad:},
\end{align}
where
\begin{subequations}
\begin{align}
& \braket{:\oFz_c(\omega)\oFzd_c:}
\nonumber \\ &
= \Gc(\omega)\{ \braket{\oad_{\low}\oa_{\low}}+\braket{\oa_{\rai}\oad_{\rai}} \}
+ \Gc(\omega)_+\{ \braket{\oad_{\low}\oa_{\rai}} + \braket{\oa_{\low}\oad_{\rai}} \}
\nonumber \\ & \quad
+ \Gc(\omega)_+^* \{ \braket{\oa_{\rai}\oad_{\low}} + \braket{\oad_{\rai}\oa_{\low}} \},
\end{align}
\begin{align}
& \braket{:\oa(\omega)\oFzd_c:}
\nonumber \\ &
= \braket{\oFz_{c\low}(\omega)^{\dagger}\oa_{\low}}
+ \braket{\oFz_{c\low}(\omega)^{\dagger}_-\oa_{\rai}} + \braket{\oa_{\rai}(\omega)_-\oFzd_{c\low}}
\nonumber \\ & \quad
+ \braket{\oa_{\low}(\omega)_+\oFzd_{c\rai}} + \braket{\oFz_{c\rai}(\omega)_+^{\dagger}\oa_{\low}}
+ \braket{\oa_{\rai}(\omega)\oFzd_{c\rai}},
\end{align}
\begin{align}
& \braket{:\oFz_c(\omega)\oad:}
\nonumber \\ &
= \braket{\oa_{\low}(\omega)^{\dagger}\oFz_{c\low}}
+ \braket{\oa_{\low}(\omega)_-^{\dagger}\oFz_{c\rai}} + \braket{\oFz_{c\rai}(\omega)_-\oad_{\low}}
\nonumber \\ & \quad
+ \braket{\oFz_{c\low}(\omega)_+\oad_{\rai}} + \braket{\oa_{\rai}(\omega)_+^{\dagger}\oFz_{c\low}}
+ \braket{\oFz_{c\rai}(\omega)\oad_{\rai}},
\end{align}
\begin{align}
& \braket{:\oa(\omega)\oad:}
\nonumber \\ &
= \braket{\oa_{\low}(\omega)^{\dagger}\oa_{\low}}
+ \braket{\oa_{\low}(\omega)_-^{\dagger}\oa_{\rai}} + \braket{\oa_{\rai}(\omega)_-\oad_{\low}}
\nonumber \\ & \quad
+ \braket{\oa_{\low}(\omega)_+\oad_{\rai}} + \braket{\oa_{\rai}(\omega)_+^{\dagger}\oa_{\low}}
+ \braket{\oa_{\rai}(\omega)\oad_{\rai}}.
\end{align}
\end{subequations}
On the other hand, the phase-sensitive correlation is expressed as
\begin{align}
& \braket{:\oFout_c(\omega)\oFout_c:}
\nonumber \\ &
= \braket{:\oFz_c(\omega)\oFz_c:}
+ \Gc(\omega)\braket{:\oa(\omega)\oFz_c:}
\nonumber \\ & \quad
+ \Gc(-\omega)\braket{:\oFz_c(\omega)\oa:}
+ \Gc(\omega)\Gc(-\omega)\braket{:\oa(\omega)\oa:},
\end{align}
\begin{subequations}
\begin{align}
& \braket{:\oFz_c(\omega)\oFz_c:} \nonumber \\
& = \{ \Gc(\omega)_+ + \Gc(-\omega)_+ \}
  \{ 2\braket{\oa_{\rai}\oa_{\low}} + \braket{\oa_{\rai}\oa_{\rai}} + \braket{\oa_{\low}\oa_{\low}} \},
\end{align}
\begin{align}
& \braket{:\oa(\omega)\oFz_c:}
\nonumber \\ &
= \braket{\oa_{\low}(\omega)_+\oFz_{c\low}} + \braket{\oFz_{c\low}(-\omega)_+\oa_{\low}}
+ \braket{\oFz_{c\rai}\oa_{\low}(\omega)}
\nonumber \\ & \quad
+ \braket{\oa_{\rai}(\omega)\oFz_{c\low}}
+ \braket{\oFz_{c\rai}(-\omega)_-\oa_{\rai}} + \braket{\oa_{\rai}(\omega)_-\oFz_{c\rai}},
\end{align}
\begin{align}
& \braket{:\oFz_c(\omega)\oa:}
\nonumber \\ &
= \braket{\oFz_{c\low}(\omega)_+\oa_{\low}} + \braket{\oa_{\low}(-\omega)_+\oFz_{c\low}}
+ \braket{\oa_{\rai}\oFz_{c\low}(\omega)}
\nonumber \\ & \quad
+ \braket{\oFz_{c\rai}(\omega)\oa_{\low}}
+ \braket{\oa_{\rai}(-\omega)_-\oFz_{c\rai}} + \braket{\oFz_{c\rai}(\omega)_-\oa_{\rai}},
\end{align}
\begin{align}
& \braket{:\oa(\omega)\oa:}
\nonumber \\ &
= \braket{\oa_{\low}(\omega)_+\oa_{\low}} + \braket{\oa_{\low}(-\omega)_+\oa_{\low}}
+ \braket{\oa_{\rai}(-\omega)\oa_{\low}}
\nonumber \\ & \quad
+ \braket{\oa_{\rai}(\omega) \oa_{\low}}
+ \braket{\oa_{\rai}(-\omega)_-\oa_{\rai}} + \braket{\oa_{\rai}(\omega)_-\oa_{\rai}}.
\end{align}
\end{subequations}
We have numerically checked that the correlation functions
$\braket{:\oFout_c(\omega)\oFoutd_c:}$ and
$\braket{:\oFout_c(\omega)\oFout_c:}$
are approximately zero
if the $\oHz$ system is in the ground state $\orhoss = \ketg\brag$.
The small deviation comes from the approximation in which
the density operator $\orho(t)$ is moved outside the time integral
in the master equation \eqref{eq:master_Born}.

\section{Calculation of observables in input-output formalism} \label{app:ordering_input-output}
Let's calculate the emission and squeezing of photonic output
from the cavity in the input-output formalism.
The coefficient matrix \eqref{eq:mM}
of the Langevin equations is diagonalized as
\begin{equation}
\mM(\omega) = \mV(\omega)
\bm{\mathsf{D}}\begin{bmatrix}
\wn_L(\omega)\\
\wn_U(\omega) \\
-\wn_L(-\omega)^* \\
-\wn_U(-\omega)^*
\end{bmatrix}
\mV(\omega)^{-1},
\end{equation}
where $\bm{\mathsf{D}}[\cdots]$ represents an diagonal matrix
with elements $\cdots$.
Due to the coupling with reservoirs, the eigen values $\{\wn_{L,U}(\omega)\}$
depend on frequency $\omega$, and
are modified from the original eigen frequencies $\{\omega_{L,U}\}$
derived from Eq.~\eqref{eq:eigen_state_closed}.
The modification depends on the strengths $\kappa_j$ and $\gamma_j$
of system-reservoir coupling.
We redefine Langevin (fluctuation) operators
in the polariton basis as
\begin{equation} \label{eq:FLU=invVFcx} 
\ovFnz(\omega) = 
\begin{bmatrix}
\oFnz_L(\omega) \\ \oFnz_U(\omega) \\ \oFnz_L(-\omega)^{\dagger} \\ \oFnz_U(-\omega)^{\dagger}
\end{bmatrix}
= \mV(\omega)^{-1}
\begin{bmatrix}
\oFz_c(\omega) \\ \oFz_x(\omega) \\ \oFz_c(-\omega)^{\dagger} \\ \oFz_x(-\omega)^{\dagger}
\end{bmatrix}.
\end{equation}
Because of the modification of the coefficients,
these operators are in general different from
the Fourier transform of free field of the reservoir field $\oF_{L,U}$,
Eq.~\eqref{eq:FLU=Fc+Fx+Fcd+Fxd}, in the polariton basis.
However, if the system-reservoir coupling is weak enough
compared to the characteristic frequency of the polariton system,
the redefined operators are approximately equal to 
the Fourier transform of Eq.~\eqref{eq:oFzLU}.
At the same time, the Born approximation used in the master equation
is also valid.
The photon and excitation operators are then represented
in the frequency domain as
\begin{equation} \label{eq:a(w)=L(w)F(w)} 
\begin{bmatrix}
\oa(\omega) \\ \ob(\omega) \\ \oa(-\omega)^{\dagger} \\ \ob(-\omega)^{\dagger}
\end{bmatrix}
= \mL(\omega) \ovFnz(\omega),
\end{equation}
where
\begin{equation}
\mL(\omega)
= [\mM(\omega)-\omega\munit]^{-1} \mV(\omega).
\end{equation}
Substituting Eqs.~\eqref{eq:FLU=invVFcx} and \eqref{eq:a(w)=L(w)F(w)} into it,
the photonic output operator is represented as
\begin{align}
\oFout_c(\omega)
& =   T_L(\omega) \oFnz_L(\omega)
  + T_U(\omega) \oFnz_U(\omega)
\nonumber \\ & \quad
  + S_L(\omega) \oFnz_L(-\omega)^{\dagger}
  + S_U(\omega) \oFnz_U(-\omega)^{\dagger},
 \label{eq:oFout_c} 
\end{align}
where
\begin{subequations}
\begin{align}
T_L(\omega) & = V_{11}(\omega)+\Gc(\omega)L_{11}(\omega), \\
T_U(\omega) & = V_{12}(\omega)+\Gc(\omega)L_{12}(\omega), \\
S_L(\omega) & = V_{13}(\omega)+\Gc(\omega)L_{13}(\omega), \\
S_U(\omega) & = V_{14}(\omega)+\Gc(\omega)L_{14}(\omega).
\end{align}
\end{subequations}

As discussed in Sec.~\ref{sec:Langevin},
for describing the dissipation of $\oHz$ system,
we consider the correlation of free fields as in Eq.~\eqref{eq:<Fzcx(w)Fzcx>}.
Precisely speaking, in order to guarantee no photon detection,
the expectation values in the ground state should be replaced
by
\begin{equation}
\begin{pmatrix}
\braketg{\oa\oad} & \braketg{\oa\obd} & \braketg{\oa\oa} & \braketg{\oa\ob} \\
\braketg{\ob\oad} & \braketg{\ob\obd} & \braketg{\ob\oa} & \braketg{\ob\ob} \\
\braketg{\oad\oad} & \braketg{\oad\obd} & \braketg{\oad\oa} & \braketg{\oad\ob} \\
\braketg{\obd\oad} & \braketg{\obd\obd} & \braketg{\obd\oa} & \braketg{\obd\ob}
\end{pmatrix}
\rightarrow\mV(\omega)
\begin{pmatrix}
1 & 0 & 0 & 0 \\
0 & 1 & 0 & 0 \\
0 & 0 & 0 & 0 \\
0 & 0 & 0 & 0
\end{pmatrix}
\mV(\omega)^{*\text{T}},
\end{equation}
because of the modification of coefficients $\mV(\omega)$.
If we detect the output photons outside the cavity, the detection process should be dissipative
for the whole system,
and virtual photons should not be counted.
Then, when we calculate emission spectrum
$\braket{:\oFout_c(\omega)\oFoutd_c:}$
and phase-sensitive correlation $\braket{:\oFout_c(\omega)\oFout_c:}$,
the fluctuation (Langevin) operators $\{\oFz_{L,U}(\omega)\}$
should be normal- and time-ordered
obeying the theory of measurement \cite{Carmichael1987JOSAB,gardiner04,walls08}.
Then, the correlation functions of output photonic field \eqref{eq:oFout_c}
are represented as
\begin{widetext}
\begin{multline}
\braket{:\oFout_c(\omega)\oFoutd_c:} \\
= \sum_{j,k=L,U} \left\{
    T_j(\omega)\braket{\oFnz_k(\omega)^{\dagger}\oFnz_j} T_k(\omega)^*
  + T_j(\omega)\left[
      \braket{\oFnz_j(\omega)_+\oFnz_k}
    + \braket{\oFnz_k(-\omega)_+\oFnz_j}
    \right] S_k(\omega)^*
\right. \\ \left.
  + S_j(\omega) \left[
      \braket{\oFnz_k(-\omega)_-^{\dagger}\oFnzd_j}
    + \braket{\oFnz_j(\omega)_-^{\dagger}\oFnzd_k}
    \right] T_k(\omega)^*
  + S_j(\omega) \braket{\oFnz_j(-\omega)^{\dagger}\oFnz_k} S_k(\omega)^*
  \right\},
\end{multline}
\begin{multline}
\braket{:\oFout_c(\omega)\oFout_c:} \\
= \sum_{j,k=L,U} \left\{
    T_j(\omega)\left[
      \braket{\oFnz_j(\omega)_+\oFnz_k}
    + \braket{\oFnz_k(-\omega)_+\oFnz_j}
    \right] T_k(-\omega)
  + T_j(\omega)\braket{\oFnz_k(\omega)^{\dagger}\oFnz_j} S_k(-\omega)
\right. \\ \left.
  + S_j(\omega)\braket{\oFnz_j(-\omega)^{\dagger}\oFnz_k} T_k(-\omega)
  + S_j(\omega)\left[
      \braket{\oFnz_k(-\omega)_-^{\dagger}\oFnzd_j}
    + \braket{\oFnz_j(\omega)_-^{\dagger}\oFnzd_k}
    \right] S_k(-\omega)
  \right\}.
\end{multline}
\end{widetext}
If we consider the fluctuation correlation as shown
in Eq.~\eqref{eq:<Fzcx(w)Fzcx>},
we can numerically verify that both of them are completely zero.


\end{document}